\documentclass[aps]{revtex4}  
\usepackage{graphicx}  
\usepackage{dcolumn}   
\usepackage{bm}        
\usepackage{amssymb}   
\usepackage{rotating}  
\def\MET{{\mbox{$E\kern-0.57em\raise0.19ex\hbox{/}_{T}$}}}
\def\met{{\mbox{$E\kern-0.57em\raise0.19ex\hbox{/}_{T}$}}}
\def\DZ{D\O\ }
\def\DZero{D\O\ }
\def\Dzero{D\O\ }

\def\ifb{~fb$^{-1}$}
\def\pp{$p\bar{p}$}

\def\ttbar{$t\bar{t}$}
\def\WH{$WH\rightarrow \ell\nu b\bar{b}$}
\def\WHt{$WH\rightarrow \tau\nu b\bar{b}$}
\def\lmet{$WH\rightarrow \ell\kern-0.45em\raise0.19ex\hbox{/} \nu b\bar{b}$}
\def\ZH{$ZH\rightarrow \nu\bar{\nu} b\bar{b}$}
\def\ZHll{$ZH\rightarrow \ell^+ \ell^- b\bar{b}$}

\def\www{$WH \rightarrow WW^{+} W^{-}$}

\def\hww{$H\rightarrow W^+ W^-$}
\def\hbb{$H\rightarrow b\bar{b}$}

\def\tevE{$\sqrt{s}=1.96$~TeV}

\begin{document}
\rightline{FERMILAB-PUB-09-060-E}
\rightline{CDF Note 9713}
\rightline{\DZ Note 5889}
\vskip0.5in

\title{Combined CDF and \DZ Upper Limits on Standard Model Higgs-Boson Production with up to 4.2 fb$^{-1}$ of Data\\[2.5cm]}

\author{
The TEVNPH Working Group\footnote{The Tevatron
New-Phenomena and Higgs working group can be contacted at
TEVNPHWG@fnal.gov. More information can be found at http://tevnphwg.fnal.gov/.}
 } 
\affiliation{\vskip0.3cm for the CDF and \DZ Collaborations\\ \vskip0.2cm
\today} 
\begin{abstract} 
\vskip0.3in 
We combine results from CDF and D\O\ on direct searches for a standard
model (SM) Higgs boson ($H$) in \pp~collisions at the Fermilab
Tevatron at $\sqrt{s}=1.96$~TeV. Compared to the previous Higgs
Tevatron combination, more data and new channels ($WH \rightarrow \tau
\nu b \bar{b}$, $VH \rightarrow \tau \tau b \bar{b}/jj \tau\tau$, $VH
\rightarrow jj b \bar{b}$, $t \bar{t} H \rightarrow t \bar{t} b
\bar{b}$) have been added.  Most previously used channels have been
reanalyzed to gain sensitivity. We use the latest parton distribution
functions and $gg \rightarrow H$ theoretical cross sections when
comparing our limits to the SM predictions. 
 With 2.0-3.6\ifb\ of data
analyzed at CDF, and 0.9-4.2\ifb\ at D\O, the 95\% C.L. upper limits
on Higgs boson production are a factor of 2.5~(0.86) times the SM
cross section for a Higgs boson mass of $m_{H}=$115~(165)~GeV/c$^2$.
Based on simulation, the corresponding median expected upper limits
are 2.4 (1.1). 
The mass range excluded at 95\% C.L. for a SM
Higgs has been extended to  $160<m_{H}<170$~GeV/c$^{2}$.
\\[2cm]
{\hspace*{5.5cm}\em Preliminary Results}
\end{abstract}

\maketitle

\newpage
\section{Introduction} 

The search for a mechanism for electroweak symmetry breaking, and in
particular for a standard model (SM) Higgs boson has been a major goal
of particle physics for many years, and is a central part of the
Fermilab Tevatron physics program. Both the CDF and \Dzero experiments
are reporting new combinations~\cite{CDFhiggs,DZhiggs} of multiple
direct searches for the SM Higgs boson.  The new searches include more
data and improved analysis techniques compared to previous analyses.
The sensitivities of these new combinations significantly exceed
previous work~\cite{CDFHiggsICHEP,DZHiggsICHEP}.  The most recent
Tevatron Higgs combination~\cite{TevHiggsICHEP} only included channels
seeking Higgs bosons of masses between 155 and 200~GeV/$c^2$, and the
most recent combination over the entire mass range 100-200~GeV/$c^2$
was reported in April 2008~\cite{tev-apr-2008}.

In this note, we combine the most recent results of all such searches
in \pp~collisions at~\tevE.  The analyses combined here seek signals
of Higgs bosons produced in associated with vector bosons
($q\bar{q}\rightarrow W/ZH$), through gluon-gluon fusion
($gg\rightarrow H$), and through vector boson fusion (VBF) ($q\bar{q}\rightarrow q^{\prime}\bar{q}^{\prime}H$)
corresponding to integrated luminosities ranging from 2.0-3.6\ifb~at
CDF and 0.9-4.2\ifb~at D\O. The Higgs boson decay modes studied are
$H\rightarrow b{\bar{b}}$, $H\rightarrow W^+W^-$, $H\rightarrow
\tau^+\tau^-$ and $H\rightarrow \gamma\gamma$.

To simplify the combination, the searches are separated into 75
mutually exclusive final states (23 for CDF and 52 for D\O; see
Table~\ref{tab:cdfacc} and ~\ref{tab:dzacc}) referred to as
``analyses'' in this note.  The selection procedures for each analysis
are detailed in Refs.~\cite{cdfWH} through~\cite{dzttH}, and are
briefly described below.

\section{Acceptance, Backgrounds and Luminosity}  

Event selections are similar for the corresponding CDF and D\O\ analyses.
For the case of \WH, an isolated lepton ($\ell=$ electron or muon) and 
two jets are required, with one or more $b$-tagged jet, i.e., identified 
as containing a weakly-decaying $B$ hadron.  
Selected events must also display a significant imbalance 
in transverse momentum
(referred to as missing transverse energy or \met).  Events with more than one
isolated lepton are vetoed.  
For the D\O\ \WH\ analyses, two and three jet events are analyzed separately, and 
in each of these samples
two non-overlapping $b$-tagged samples are
defined, one being a single ``tight'' $b$-tag (ST) sample, and the other a 
double ``loose'' $b$-tag (DT) sample. The tight and loose $b$-tagging criteria
are defined with respect to the mis-identification 
rate that the $b$-tagging algorithm yields for light quark or gluon jets 
(``mistag rate'') typically $\le 
0.5\%$ or $\le 1.5\%$, respectively.  
The final variable is a neural network output which takes as input seven kinematics
variables and a matrix element discriminant for the 2 jet sample, while for the
3 jet sample the dijet invariant mass is used.
In this combination, we add a new analysis
\WHt\ in which the $\tau$ is identified through its hadronic decays.
this analysis is sensitive to $ZH \rightarrow \tau \met b \bar{b}$
 as well in those cases where a $\tau$ fails to be identified.
The analysis is carried out according to the type of reconstructed 
$\tau$ and  is also 
separated into two and three jets with DT events only.
It uses the dijet invariant
mass of the $b\bar{b}$ system as discriminant variable.

For the CDF \WH\ analyses, the events are grouped into six categories.  In addition
to the selections requiring an identified lepton, events with an isolated track
failing lepton selection requirements are grouped into their own categories.  
This provides some acceptance for single prong tau decays.
Within the
lepton categories there 
are three $b$-tagging categories -- two tight $b$-tags (TDT), one tight $b$-tag
and one loose $b$-tag (LDT), and a single, tight, $b$-tag (ST).  In each category, two discriminants
are calculated for each event.  One neural network discriminant is trained at each
$m_H$ in the test range, separately for each category.  A second discriminant is a 
boosted decision tree, featuring not only event kinematic and $b$-tagging observables,
but matrix element discriminants as well.  These two discriminants are then combined
together using an evolutionary neural network~\cite{NEAT} to form a single discriminant with optimal
performance.

For the \ZH\ analyses, the selection is
similar to the $WH$ selection, except all events with isolated leptons are vetoed and
stronger multijet background suppression techniques are applied. 
Both CDF and D\O\  analyses use a track-based missing transverse momentum calculation
as a discriminant against false \met .  
There is a sizable fraction of \WH\ signal in which the lepton is undetected,
that is selected in the \ZH\ samples,  so these analyses are also refered to as
$VH \rightarrow \met b \bar{b}$.
The CDF  analysis uses three non-overlapping samples of 
events (TDT, LDT and ST as for $WH$) while D\O\
uses a sample of events having one tight $b$-tag jet and one loose $b$-tag jet.
CDF used
neural-network discriminants as the final variables, while D\O\ 
uses boosted decision trees as advanced analysis technique.

The \ZHll\ analyses require two isolated leptons and
at least two jets. They   use non-overlapping samples of
events with one tight $b$-tag and two loose $b$-tags. 
For the D\O\ analysis  neural-network and boosted decision trees
discriminants are the 
final variables for setting  limits (depending on the sub-channel), 
while  CDF uses the output
of a 2-dimensional neural-network.  
CDF corrects jet energies for \met\ using a neural network approach.
In this analysis  also the events are
divided  into three tagging
categories: tight double tags, loose double tags, and single tags.

For the \hww~analyses, signal events are characterized by
a large \met~and two opposite-signed, isolated leptons.  The presence of
neutrinos in the final state prevents the reconstruction of the
candidate Higgs boson mass. 
 D\O\ selects events containing electrons and muons,
dividing the data sample into three final states:
$e^+e^-$, $e^\pm \mu^\mp$, and $\mu^+\mu^-$.
CDF separates the \hww\ events in five non-overlapping samples,
labeled ``high $s/b$'' and ``low $s/b$'' for the lepton selection categories, and
also split by the number of jets: 0, 1, or 2+ jets.  The sample with two or more jets
is not split into low $s/b$ and high $s/b$ lepton categories.   The division of events
into jet categories allows the analysis discriminants to separate three different categories of
signals from the backgrounds more effectively.  The signal production mechanisms considered are
$gg\rightarrow H\rightarrow W^+W^-$, $WH+ZH\rightarrow jjW^+W^-$, and the vector-boson
fusion process.  The final discriminants are neural-network outputs for D\O\, and
neural-network output
including likelihoods constructed from matrix-element probabilities (ME) as input
to the neural network, for CDF, in the 0-jet bin, else the ME are not used.
 All analyses in this channel have
been updated with more data and analysis improvements.

The CDF collaboration also contributes an analysis searching for Higgs bosons decaying
to a tau lepton pair, in three separate production channels:
direct $p \bar{p} \rightarrow H$ production, associated $WH$ or $ZH$ production,
or vector boson production with $H$ and forward jets in the final state.
In  this analysis, the final variable for setting  limits is
a combination of several neural-network discriminants.

D\O\ also
contributes a new analysis 
for the final state $\tau
\tau$ jet jet, which is sensitive to the $VH\rightarrow jj \tau \tau$, $ZH 
\rightarrow \tau \tau b \bar{b}$, VBF and gluon gluon fusion
(with two additional jets) mechanisms. It
uses a neural network output  as discriminant variable. 

The CDF collaboration introduces a new all-hadronic channel, $WH+ZH\rightarrow jjb{\bar{b}}$ for this
combination.  Events are selected with four jets, at least two of which are $b$-tagged
with the tight $b$-tagger.  The large QCD backgrounds are estimated with the use of
data control samples, and the final variable is a matrix element signal probability
discriminant.

The D\O\ collaboration
contributes three \www\ analyses, where the associated $W$ boson and
the $W$ boson from the Higgs boson decay which has the same charge are required 
to decay leptonically,
thereby defining three like-sign dilepton final states 
($e^\pm e^\pm$, $e^\pm \mu^\pm$, and $\mu^{\pm}\mu^{\pm}$)
containing all decays of the third
$W$ boson. In  this analysis, which has not been updated for this combination,
the final variable is a likelihood discriminant formed from several
topological variables.   CDF contributes a \www\ analysis using a selection of like-sign
dileptons and a neural network to further purify the signal.
D\O\ also contributes an analysis searching for direct Higgs boson production
decaying to a photon pair in 4.2 fb$^{-1}$ of data.
In  this analysis, the final variable is the invariant mass of the two-photon system.

Another new search from D\O\ is
included in this combination, namely the search for
$t \bar{t} H \rightarrow t \bar{t} b \bar{b}$. Here the samples are analyzed 
independently according to the
number of $b$-tagged jets (1,2,3, i.e. ST,DT,TT) 
and the total number of jets (4 or 5).
The total transverse energy of the reconstructed objects ($H_T$) is used
as discriminant variable.

All Higgs boson signals are simulated using \textsc{PYTHIA}~\cite{pythia}, and 
\textsc{CTEQ5L} or \textsc{CTEQ6L}~\cite{cteq} leading-order (LO)
parton distribution functions.  
The $gg\rightarrow H$ production cross section is calculated at NNLL in
QCD and also includes two-loop electroweak
effects; see Refs.~\cite{anastasiou,grazzinideflorian} and references therein for 
the different steps of these calculations.   The newer calculation
includes a more thorough treatment of higher-order radiative corrections, particularly those involving
$b$ quark loops.  The $gg\rightarrow H$ production cross section depends strongly on the PDF set chosen and the
accompanying value of $\alpha_s$.  The cross sections used 
here are calculated with the MSTW 2008 NNLO PDF set~\cite{mstw2008}.   The new $gg\rightarrow H$ cross sections
supersede those used in the update of Summer 2008~\cite{TevHiggsICHEP,nnlo1,aglietti}, which had a simpler
treatment of radiative corrections and used the older MRST 2002 PDF set~\cite{mrst2002}.  The Higgs boson
production cross sections used here
are listed in Table~\ref{tab:higgsxsec}~\cite{grazzinideflorian}.  
We include all significant Higgs production modes in the high mass search: besides gluon-gluon 
fusion through a virtual top quark loop  (ggH), 
we include production in association 
with a $W$ or $Z$ vector boson  (VH)~\cite{nnlo2,Brein,Ciccolini}, 
and vector boson  fusion (VBF)~\cite{nnlo2,Berger}.

The Higgs boson
decay branching ratio predictions are calculated with HDECAY~\cite{hdecay}.
For both CDF and D\O , events from
multijet (instrumental) backgrounds (``QCD production'') are measured
in data with different methods, in orthogonal samples.
For CDF, backgrounds
from other SM processes were generated using \textsc{PYTHIA},
\textsc{ALPGEN}~\cite{alpgen}, \textsc{MC@NLO}~\cite{MC@NLO}
 and \textsc{HERWIG}~\cite{herwig}
programs. For D\O , these backgrounds were generated using
\textsc{PYTHIA}, \textsc{ALPGEN}, and \textsc{COMPHEP}~\cite{comphep},
with \textsc{PYTHIA} providing parton-showering and hadronization for
all the generators.  These background processes were normalized using either
experimental data or next-to-leading order calculations (from
\textsc{MCFM}~\cite{mcfm} for $W+$ heavy flavor process).

Integrated luminosities, 
and references to the collaborations' public documentation for each analysis
are given in Table~\ref{tab:cdfacc}
for CDF and in Table~\ref{tab:dzacc} for D\O .  
The tables include the ranges of Higgs boson mass ($m_H$) over which
the searches were performed. 


\begin{table}[h]
\caption{\label{tab:cdfacc}Luminosity, explored mass range and references
for the different processes
and final state ($\ell=e, \mu$) for the CDF analyses
}
\begin{ruledtabular}
\begin{tabular}{lccc} \\
Channel & Luminosity (fb$^{-1}$) & $m_H$ range (GeV/$c^2$) & Reference \\ \hline
$WH\rightarrow \ell\nu b\bar{b}$ \ \ \ 2$\times$(TDT,LDT,ST)        & 2.7  & 100-150 & \cite{cdfWH} \\
$ZH\rightarrow \nu\bar{\nu} b\bar{b}$ \ \ \ (TDT,LDT,ST)            & 2.1  & 105-150 & \cite{cdfZH} \\
$ZH\rightarrow \ell^+\ell^- b\bar{b}$ \ \ \ 2$\times$(TDT,LDT,ST)   & 2.7  & 100-150 & \cite{cdfZHll} \\
$H\rightarrow W^+ W^-$ \ \ \ (low,high $s/b$)$\times$(0,1 jets)+(2+ jets)  & 3.6  & 110-200 & \cite{cdfHWW} \\
$WH \rightarrow WW^+ W^- \rightarrow \ell^\pm\nu \ell^\pm\nu$ & 3.6  & 110-200 & \cite{cdfHWW} \\
$H$ + $X\rightarrow \tau^+ \tau^- $ + 2 jets                  & 2.0  & 110-150 & \cite{cdfHtt} \\
$WH+ZH\rightarrow jjb{\bar{b}}$                               & 2.0  & 100-150 & \cite{cdfjjbb} \\ 
\end{tabular}
\end{ruledtabular}
\end{table}

\vglue 0.5cm 

\begin{table}[h]
\caption{\label{tab:dzacc}Luminosity, explored mass range and references 
for the different processes
and final state ($\ell=e, \mu$) for the D\O\ analyses
}
\begin{ruledtabular}
\begin{tabular}{lccc} \\
Channel & Luminosity (fb$^{-1}$) & $m_H$ range (GeV/$c^2$) & Reference \\ \hline
$WH\rightarrow \ell\nu b\bar{b}$ \ \ \ 2$\times$(ST,DT)             & 2.7  & 100-150 & \cite{dzWHl} \\
$WH\rightarrow \tau\nu b\bar{b}$ \ \ \ 2$\times$(ST,DT)             & 0.9  & 105-145 & \cite{dzWHt} \\
$VH\rightarrow \tau\tau b\bar{b}/q\bar{q} \tau\tau$           & 1.0  & 105-145 & \cite{dzWHt} \\
$ZH\rightarrow \nu\bar{\nu} b\bar{b}$ \ \ \ (DT)                    & 2.1  & 105-145 & \cite{dzZHv} \\
$ZH\rightarrow \ell^+\ell^- b\bar{b}$ \ \ \ 2$\times$(ST,DT)        & 2.3  & 105-145 & \cite{dzZHll} \\
$WH \rightarrow WW^+ W^- \rightarrow \ell^\pm\nu \ell^\pm\nu$ & 1.1  & 120-200 & \cite{dzWWW} \\
$H\rightarrow W^+ W^- \rightarrow \ell^\pm\nu \ell^\mp\nu$    & 3.0--4.2  & 115-200 & \cite{dzHWW}\\
$H \rightarrow \gamma \gamma$                                 & 4.2  & 100-150 & \cite{dzHgg} \\ 
$t \bar{t} H \rightarrow t \bar{t} b \bar{b}$ \ \ \ 2$\times$(ST,DT,TT)& 2.1  & 105-145 & \cite{dzttH} \\ 
\end{tabular}
\end{ruledtabular}
\end{table}

\begin{table}
\caption{
The (N)NLO production cross sections and decay branching fractions for the SM
Higgs boson assumed for the combination}
\vspace{0.2cm}
\label{tab:higgsxsec}
\begin{ruledtabular}
\begin{tabular}{cccccccc}\\
$m_H$ & $\sigma_{gg\rightarrow H}$ & $\sigma_{WH}$ & $\sigma_{ZH}$ & $\sigma_{VBF}$ &  
$B(H\rightarrow b{\bar{b}})$ & $B(H\rightarrow \tau^+{\tau^-})$ & $B(H\rightarrow W^+W^-)$ \\ 
(GeV/$c^2$) & (fb) & (fb) & (fb) & (fb) & (\%) & (\%) & (\%) \\ \hline
   100 &    1861   &   286.1  &   166.7  &   99.5  &  81.21 & 7.924 &  1.009 \\ 
   105 &    1618   &   244.6  &   144.0  &   93.3  &  79.57 & 7.838 &  2.216 \\
   110 &    1413   &   209.2  &   124.3  &   87.1  &  77.02 & 7.656 &  4.411  \\
   115 &    1240   &   178.8  &   107.4  &   79.07 &  73.22 & 7.340 &  7.974 \\
   120 &    1093   &   152.9  &   92.7   &   71.65 &  67.89 & 6.861 &  13.20 \\
   125 &    967    &   132.4  &   81.1   &   67.37 &  60.97 & 6.210 &  20.18 \\
   130 &    858    &   114.7  &   70.9   &   62.5  &  52.71 & 5.408 &  28.69 \\
   135 &    764    &   99.3   &   62.0   &   57.65 &  43.62 & 4.507 &  38.28 \\
   140 &    682    &   86.0   &   54.2   &   52.59 &  34.36 & 3.574 &  48.33 \\
   145 &    611    &   75.3   &   48.0   &   49.15 &  25.56 & 2.676 &  58.33 \\
   150 &    548    &   66.0   &   42.5   &   45.67 &  17.57 & 1.851 &  68.17 \\
   155 &    492    &   57.8   &   37.6   &   42.19 &  10.49 & 1.112 &  78.23 \\
   160 &    439    &   50.7   &   33.3   &   38.59 &  4.00  & 0.426 &  90.11 \\
   165 &    389    &   44.4   &   29.5   &   36.09 &  1.265 & 0.136 &  96.10 \\
   170 &    349    &   38.9   &   26.1   &   33.58 &  0.846 & 0.091 &  96.53 \\
   175 &    314    &   34.6   &   23.3   &   31.11 &  0.663 & 0.072 &  95.94 \\
   180 &    283    &   30.7   &   20.8   &   28.57 &  0.541 & 0.059 &  93.45 \\
   185 &    255    &   27.3   &   18.6   &   26.81 &  0.420 & 0.046 &  83.79 \\
   190 &    231    &   24.3   &   16.6   &   24.88 &  0.342 & 0.038 &  77.61 \\
   195 &    210    &   21.7   &   15.0   &   23           &  0.295 & 0.033 &  74.95 \\
   200 &    192    &   19.3   &   13.5   &   21.19 &  0.260 & 0.029 &  73.47 \\
\end{tabular}
\end{ruledtabular}
\end{table}

\section{Distributions of Candidates} 

The number of channels combined is quite large, and the number of bins
in each channel is large.  Therefore, the task of assembling
histograms and checking whether the expected and observed limits are
consistent with the input predictions and observed data is difficult.
We therefore provide histograms that aggregate all channels' signal,
background, and data together.  In order to preserve most of the
sensitivity gain that is achieved by the analyses by binning the data
instead of collecting them all together and counting, we aggregate the
data and predictions in narrow bins of signal-to-background ratio,
$s/b$.  Data with similar $s/b$ may be added together with no loss in
sensitivity, assuming similar systematic errors on the predictions.
The aggregate histograms do not show the effects of systematic
uncertainties, but instead compare the data with the central
predictions supplied by each analysis.

The range of $s/b$ is quite large in each analysis, and so
$\log_{10}(s/b)$ is chosen as the plotting variable.  Plots of the
distributions of $\log_{10}(s/b)$ are shown for $m_H=115$ and
165~GeV/c$^2$ in Figure~\ref{fig:lnsb}.  These distributions can be
integrated from the high-$s/b$ side downwards, showing the sums of
signal, background, and data for the most pure portions of the
selection of all channels added together.  These integrals can be seen
in Figure~\ref{fig:integ}.  

 \begin{figure}[t]
 \begin{centering}
 \includegraphics[width=0.4\textwidth]{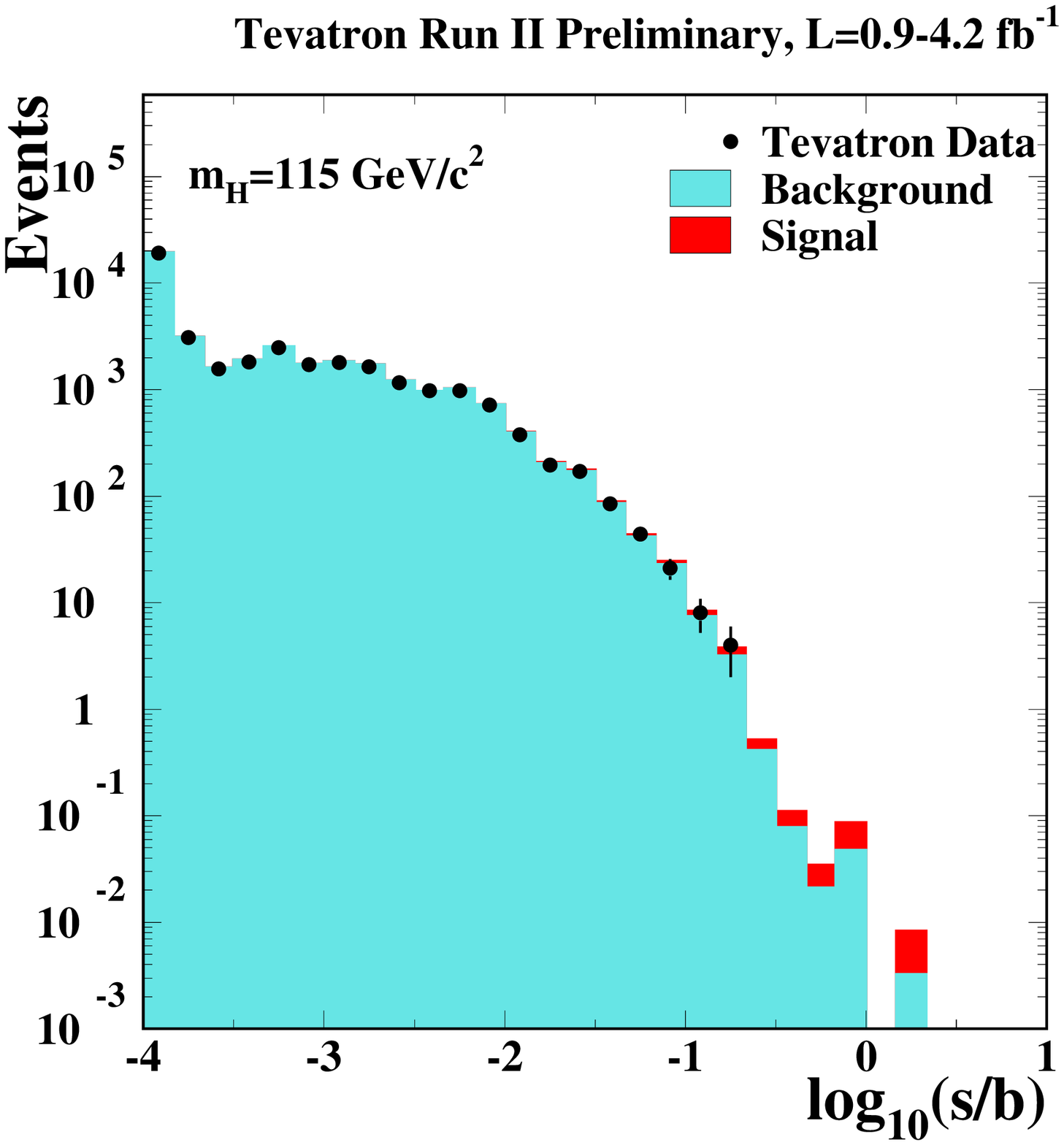}\includegraphics[width=0.4\textwidth]{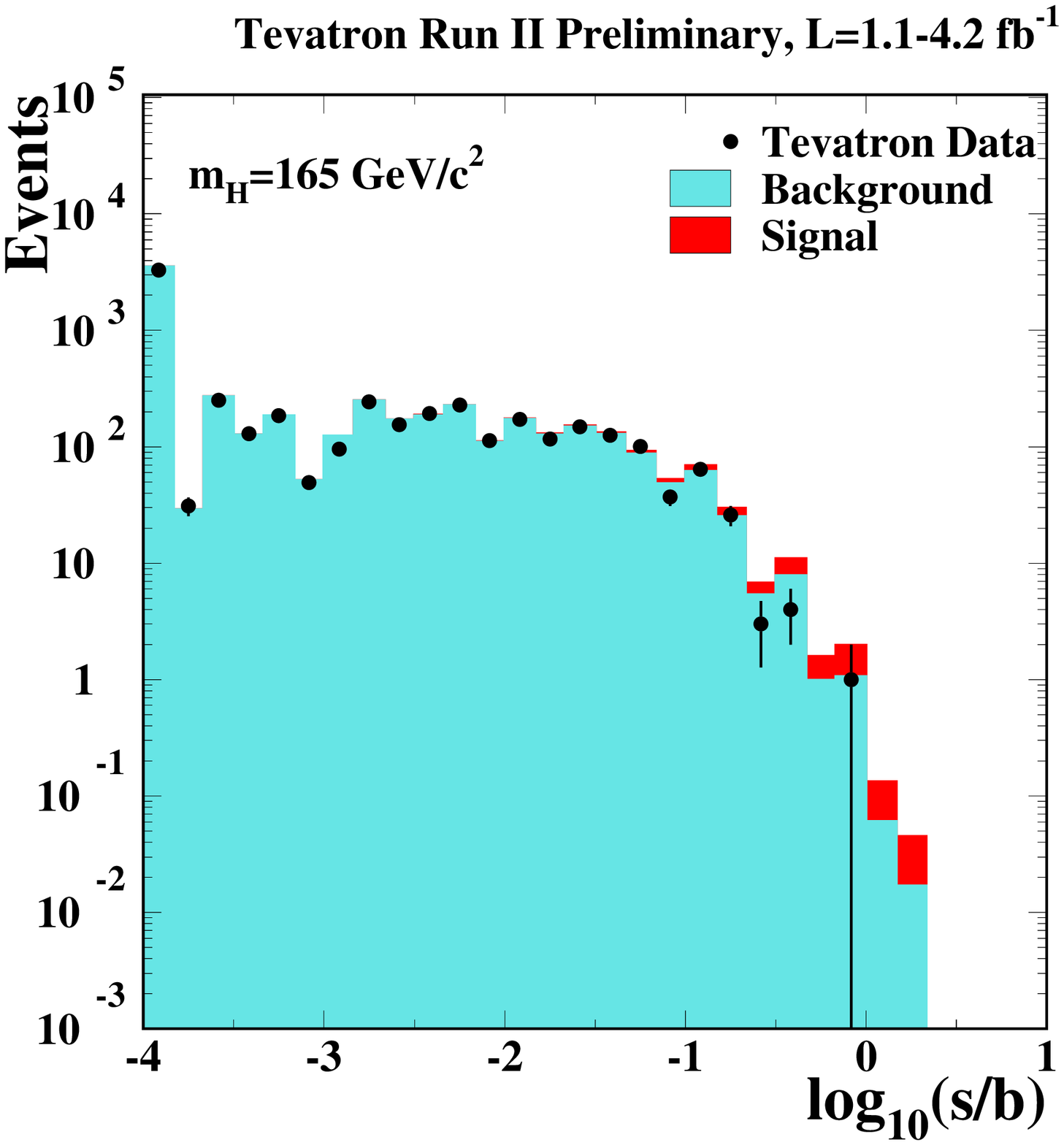}
 \caption{
 \label{fig:lnsb} Distributions of $\log_{10}(s/b)$, for the data from all contributing channels from
CDF and D\O, for Higgs boson masses of 115 and 165~GeV/$c^2$.  The
data are shown with points, and the signal is shown stacked on top of
the backgrounds.  Underflows and overflows are collected into the
bottom and top bins. }
 \end{centering}
 \end{figure}

 \begin{figure}[t]
 \begin{centering}
 \includegraphics[width=0.4\textwidth]{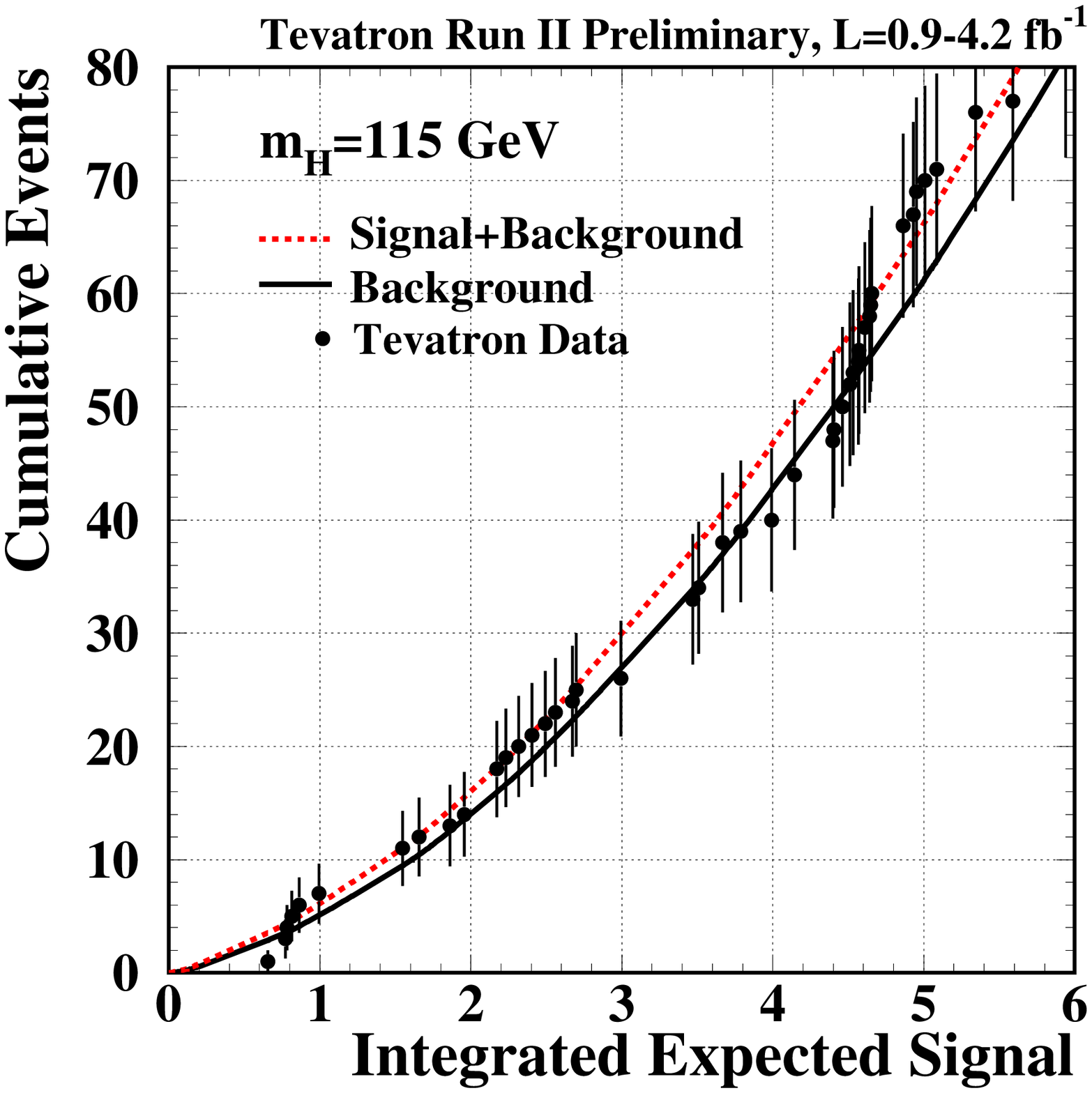}\includegraphics[width=0.4\textwidth]{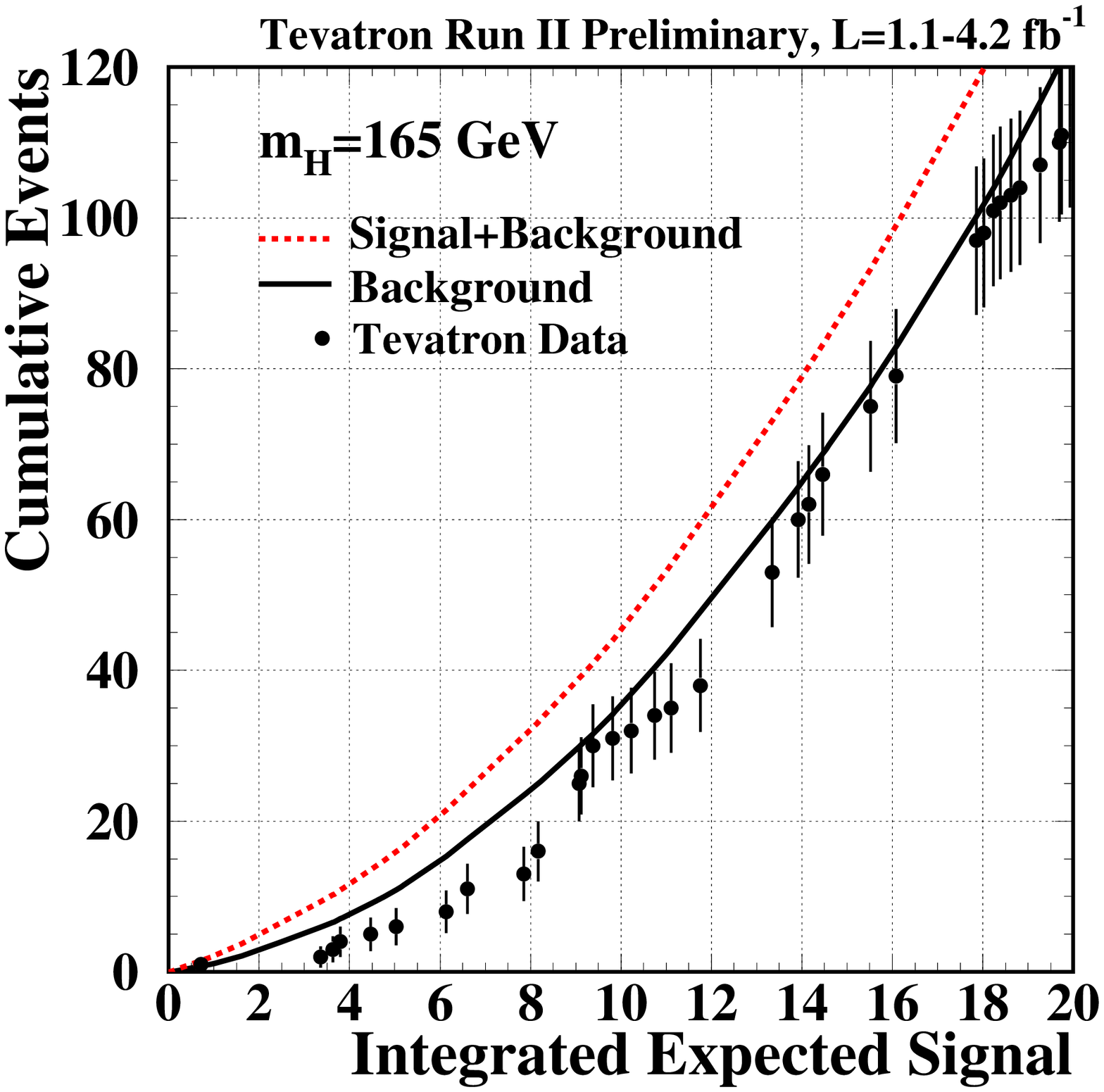}
 \caption{
 \label{fig:integ} Integrated distributions of $s/b$, starting at the high $s/b$ side.  The total signal+background
and background-only integrals are shown separately, along with the
data sums.  Data are only shown for bins that have 
data events in them.}
 \end{centering}
 \end{figure}

\section{Combining Channels} 

To gain confidence that the final result does not depend on the
details of the statistical formulation, 
we perform two types of combinations, using the
Bayesian and  Modified Frequentist approaches, which give similar results
(within 10\%).
Both methods rely on distributions in the final discriminants, and not just on
their single integrated values.  Systematic uncertainties enter as uncertainties on the
expected number of signal and background events, as well
as on the distribution of the discriminants in 
each analysis (``shape uncertainties'').
Both methods use likelihood calculations based on Poisson
probabilities.

\subsection{Bayesian Method}

Because there is no experimental information on the production cross section for
the Higgs boson, in the Bayesian technique~\cite{CDFhiggs} we assign a flat prior
for the total number of selected Higgs events.  For a given Higgs boson mass, the
combined likelihood is a product of likelihoods for the individual
channels, each of which is a product over histogram bins:

\begin{equation}
{\cal{L}}(R,{\vec{s}},{\vec{b}}|{\vec{n}},{\vec{\theta}})\times\pi({\vec{\theta}})
= \prod_{i=1}^{N_C}\prod_{j=1}^{Nbins} \mu_{ij}^{n_{ij}} e^{-\mu_{ij}}/n_{ij}!
\times\prod_{k=1}^{n_{np}}e^{-\theta_k^2/2}
\end{equation}

\noindent where the first product is over the number of channels
($N_C$), and the second product is over histogram bins containing
$n_{ij}$ events, binned in  ranges of the final discriminants used for
individual analyses, such as the dijet mass, neural-network outputs, 
or matrix-element likelihoods.
 The parameters that contribute to the
expected bin contents are $\mu_{ij} =R \times s_{ij}({\vec{\theta}}) + b_{ij}({\vec{\theta}})$ 
for the
channel $i$ and the histogram bin $j$, where $s_{ij}$ and $b_{ij}$ 
represent the expected background and signal in the bin, and $R$ is a scaling factor
applied to the signal to test the sensitivity level of the experiment.  
Truncated Gaussian priors are used for each of the nuisance parameters                                               
$\theta_k$, which define
the
sensitivity of the predicted signal and background estimates to systematic uncertainties.
These
can take the form of uncertainties on overall rates, as well as the shapes of the distributions
used for combination.   These systematic uncertainties can be far larger
than the expected SM signal, and are therefore important in the calculation of limits. 
The truncation
is applied so that no prediction of any signal or background in any bin is negative.
The posterior density function is
then integrated over all parameters (including correlations) except for $R$,
and a 95\% credibility level upper limit on $R$ is estimated
by calculating the value of $R$ that corresponds to 95\% of the area
of the resulting distribution.

\subsection{Modified Frequentist Method}

The Modified Frequentist technique relies on the $CL_s$ method, using
a log-likelihood ratio (LLR) as test statistic~\cite{DZhiggs}:
\begin{equation}
LLR = -2\ln\frac{p({\mathrm{data}}|H_1)}{p({\mathrm{data}}|H_0)},
\end{equation}
where $H_1$ denotes the test hypothesis, which admits the presence of
SM backgrounds and a Higgs boson signal, while $H_0$ is the null
hypothesis, for only SM backgrounds.  The probabilities $p$ are
computed using the best-fit values of the nuisance parameters for each
event, separately for each of the two hypotheses, and include the
Poisson probabilities of observing the data multiplied by Gaussian
constraints for the values of the nuisance parameters.  This technique
extends the LEP procedure~\cite{pdgstats} which does not involve a
fit, in order to yield better sensitivity when expected signals are
small and systematic uncertainties on backgrounds are
large~\cite{pflh}.

The $CL_s$ technique involves computing two $p$-values, $CL_{s+b}$ and $CL_b$.
The latter is defined by
\begin{equation}
1-CL_b = p(LLR\le LLR_{\mathrm{obs}} | H_0),
\end{equation}
where $LLR_{\mathrm{obs}}$ is the value of the test statistic computed for the
data. $1-CL_b$ is the probability of observing a signal-plus-background-like outcome 
without the presence of signal, i.e. the probability
that an upward fluctuation of the background provides  a signal-plus-background-like
response as observed in data.
The other $p$-value is defined by
\begin{equation}
CL_{s+b} = p(LLR\ge LLR_{\mathrm{obs}} | H_1),
\end{equation}
and this corresponds to the probability of a downward fluctuation of the sum
of signal and background in 
the data.  A small value of $CL_{s+b}$ reflects inconsistency with  $H_1$.
It is also possible to have a downward fluctuation in data even in the absence of
any signal, and a small value of $CL_{s+b}$ is possible even if the expected signal is
so small that it cannot be tested with the experiment.  To minimize the possibility
of  excluding  a signal to which there is insufficient sensitivity 
(an outcome  expected 5\% of the time at the 95\% C.L., for full coverage),
we use the quantity $CL_s=CL_{s+b}/CL_b$.  If $CL_s<0.05$ for a particular choice
of $H_1$, that hypothesis is deemed excluded at the 95\% C.L.

Systematic uncertainties are included  by fluctuating the predictions for
signal and background rates in each bin of each histogram in a correlated way when
generating the pseudoexperiments used to compute $CL_{s+b}$ and $CL_b$.

\subsection{Systematic Uncertainties} 

Systematic uncertainties differ
between experiments and analyses, and they affect the rates and shapes of the predicted
signal and background in correlated ways.  The combined results incorporate
the sensitivity of predictions to  values of nuisance parameters,
and include correlations, between rates and shapes, between signals and backgrounds,
and between channels within experiments and between experiments.
More on these issues can be found in the
individual analysis notes~\cite{cdfWH} through~\cite{dzttH}.  Here we
consider only the largest contributions and correlations between and
within the two experiments.

\subsubsection{Correlated Systematics between CDF and D\O}
The uncertainty on the measurement of the integrated luminosity is 6\%
(CDF) and 6.1\% (D\O ).  
Of this value, 4\% arises from the uncertainty
on the inelastic \pp~scattering cross section, which is correlated
between CDF and D\O . 
The uncertainty on the production rates for the signal, for 
top-quark processes (\ttbar~and single top) and for electroweak processes
($WW$, $WZ$, and $ZZ$) are taken as correlated between the two
experiments. As the methods of measuring the multijet (``QCD'')
backgrounds differ between CDF and D\O , there is no
correlation assumed between these rates.  Similarly, the large
uncertainties on the background rates for $W$+heavy flavor (HF) and $Z$+heavy flavor
are considered at this time to be uncorrelated, as both CDF and D\O\ estimate
these rates using data control samples, but employ different techniques.
The calibrations of fake leptons, unvetoed $\gamma\rightarrow e^+e^-$ conversions,
$b$-tag efficiencies and mistag rates are performed by each collaboration
using independent data samples and methods, hence are considered uncorrelated.

\subsubsection{Correlated Systematic Uncertainties for CDF}
The dominant systematic uncertainties for the CDF analyses are shown
in Table~\ref{tab:cdfsystwhtdt} 
for the $W^\pm H\rightarrow W^\pm
b{\bar{b}}$ channels, in Table~\ref{tab:cdfvvbb1}
for the $(W,Z)H\rightarrow
\MET b{\bar{b}}$ channels, in Table~\ref{tab:cdfllbb1} 
for the $ZH\rightarrow
\ell^+\ell^-b{\bar{b}}$ channels, in Table~\ref{tab:cdfsystww0} for
the $H\rightarrow W^+W^-\rightarrow \ell^{\prime \pm}\nu \ell^{\prime
\mp}\nu$ channels, in Table~\ref{tab:cdfsysttautau} for the
$H\rightarrow\tau^+\tau^-$ channel, in Table~\ref{tab:cdfjjbbsyst} for
the $WH+ZH\rightarrow jjb{\bar{b}}$ channel, and in
Table~\ref{tab:cdfsystwww} for the $WH \rightarrow WWW
\rightarrow\ell^{\prime \pm}\ell^{\prime \pm}$ channel.  Each source
induces a correlated uncertainty across all CDF channels' signal and
background contributions which are sensitive to that source.  For
\hbb, the largest uncertainties on signal arise from a scale factor
for $b$-tagging (5.3-16\%), jet energy scale (1-20\%) and MC modeling
(2-10\%).  The shape dependence of the jet energy scale, $b$-tagging
and uncertainties on gluon radiation (``ISR'' and ``FSR'') are taken
into account for some analyses (see tables).  For \hww, the largest
uncertainty comes from MC modeling (5\%).  For simulated backgrounds,
the uncertainties on the expected rates range from 11-40\% (depending
on background).  The backgrounds with the largest systematic
uncertainties are in general quite small. Such uncertainties are
constrained by fits to the nuisance parameters, and they do not affect
the result significantly.  Because the largest background
contributions are measured using data, these uncertainties are treated
as uncorrelated for the \hbb~channels. For the \hww~channel, the
uncertainty on luminosity is taken to be correlated between signal and
background. The differences in the resulting limits when treating
the remaining uncertainties as correlated or uncorrelated, is less than $5\%$.

\subsubsection{Correlated Systematic Uncertainties for D\O }
The dominant systematic uncertainties for D\O\ analyses are shown in Tables
V,\ref{tab:d0vvbb},\ref{tab:d0llbb1},\ref{tab:d0systwww},\ref{tab:d0systww},
\ref{tab:d0systgg}, and
\ref{tab:d0systtth}.
Each source induces a correlated uncertainty across all D\O\ channels
sensitive to that source.
The \hbb~analyses have an uncertainty on the
$b$-tagging rate of 3-10\% per tagged jet, and  also an
uncertainty on the jet energy and acceptance of 6-9\% (jet
identification or jet ID, energy scale, and jet resolution).
The shape dependence of 
the uncertainty on $W+$ jet modeling 
is taken into account in the limit setting, and has a small effect ($\sim 5\%$) on 
the final result.
For the \hww~and \www, the largest uncertainties are associated with lepton
measurement and acceptance. These values range from 2-11\% depending on
the final state.  The largest contributing factor to all analyses is
the uncertainty on cross sections for simulated background, and is 
6-18\%. 
All systematic uncertainties arising from the
same source are taken to be correlated between the different backgrounds and
between signal and background.

\begin{table}[t]
\caption{Systematic uncertainties on the signal contributions for
 CDF's $WH\rightarrow\ell\nu b{\bar{b}}$ tight (TDT) and loose (LDT)
 double tag, and single tag (ST) channels.  Systematic uncertainties
 are listed by name, see the original references for a detailed
 explanation of their meaning and on how they are derived.  Systematic
 uncertainties for $WH$ shown in this table are obtained for $m_H=115$
 GeV/c$^2$.  Uncertainties are relative, in percent and are symmetric
 unless otherwise indicated.  }
\label{tab:cdfsystwhtdt}
\vskip 0.2cm
{\centerline{CDF: tight double-tag (TDT) $WH\rightarrow\ell\nu b{\bar{b}}$}} 
\vskip 0.099cm
\begin{ruledtabular}
\begin{tabular}{lcccccc}\\
Contribution              & $W$+HF & Mistags & Top & Diboson & Non-$W$ & $WH$  \\ \hline
Luminosity ($\sigma_{\mathrm{inel}}(p{\bar{p}})$)          
                          & 0      & 0       & 3.8 & 3.8     & 0       &    3.8   \\
Luminosity Monitor        & 0      & 0       & 4.4 & 4.4     & 0       &    4.4   \\
Lepton ID                 & 0      & 0       & 2   & 2       & 0       &    2   \\
Jet Energy Scale          & 0      & 0       & 0   & 0       & 0       &    2   \\
Mistag Rate               & 0      & 9.0     & 0   & 0       & 0       &    0   \\
$B$-Tag Efficiency          & 0      & 0       & 8.4 & 8.4     & 0       &    8.4   \\
$t{\bar{t}}$ Cross Section  & 0    & 0       & 10  & 0       & 0       &    0   \\
Diboson Rate              & 0      & 0       & 0   & 11.5    & 0       &    0   \\
Signal Cross Section      & 0      & 0       & 0   & 0       & 0       &    5 \\
HF Fraction in W+jets     &    30.1  & 0       & 0   & 0       & 0       &    0   \\
ISR+FSR+PDF               & 0      & 0       & 0   & 0       & 0       &    5.6 \\ 
QCD Rate                  & 0      & 0       & 0   & 0       & 40      &    0   \\
\end{tabular}
\end{ruledtabular}
\vskip 0.5cm                                                                             
{\centerline{CDF: loose double-tag (LDT) $WH\rightarrow\ell\nu b{\bar{b}}$}}
\vskip 0.099cm
\label{tab:cdfsystwhldt}
\begin{ruledtabular}
\begin{tabular}{lcccccc}\\
Contribution              & $W$+HF & Mistags & Top & Diboson & Non-$W$ & $WH$  \\ \hline
Luminosity ($\sigma_{\mathrm{inel}}(p{\bar{p}})$)          
                          & 0      & 0       & 3.8 & 3.8     & 0       &    3.8   \\
Luminosity Monitor        & 0      & 0       & 4.4 & 4.4     & 0       &    4.4   \\
Lepton ID                 & 0      & 0       & 2   & 2       & 0       &    2   \\
Jet Energy Scale          & 0      & 0       & 0   & 0       & 0       &    2   \\
Mistag Rate               & 0      & 8.0     & 0   & 0       & 0       &    0   \\
$B$-Tag Efficiency          & 0      & 0       & 9.1 & 9.1     & 0       &    9.1   \\
$t{\bar{t}}$ Cross Section  & 0    & 0       & 10  & 0       & 0       &    0   \\
Diboson Rate              & 0      & 0       & 0   & 11.5    & 0       &    0   \\
Signal Cross Section      & 0      & 0       & 0   & 0       & 0       &    5 \\
HF Fraction in W+jets     &    30.1  & 0       & 0   & 0       & 0       &    0   \\
ISR+FSR+PDF               & 0      & 0       & 0   & 0       & 0       &    4.3 \\ 
QCD Rate                  & 0      & 0       & 0   & 0       & 40      &    0   \\
\end{tabular}
\end{ruledtabular}
\vskip 0.5cm                                                                             
{\centerline{CDF: single tag (ST) $WH\rightarrow\ell\nu b{\bar{b}}$}}
\vskip 0.099cm
\label{tab:cdfsystwhst}
\begin{ruledtabular}
\begin{tabular}{lcccccc}\\
Contribution              & $W$+HF & Mistags & Top & Diboson & Non-$W$ & $WH$  \\ \hline
Luminosity ($\sigma_{\mathrm{inel}}(p{\bar{p}})$)          
                          & 0      & 0       & 3.8 & 3.8     & 0       &    3.8   \\
Luminosity Monitor        & 0      & 0       & 4.4 & 4.4     & 0       &    4.4   \\
Lepton ID                 & 0      & 0       & 2   & 2       & 0       &    2   \\
Jet Energy Scale          & 0      & 0       & 0   & 0       & 0       &    2   \\
Mistag Rate               & 0      & 13.3    & 0   & 0       & 0       &    0   \\
$B$-Tag Efficiency          & 0      & 0       & 3.5 & 3.5     & 0       &    3.5   \\
$t{\bar{t}}$ Cross Section  & 0    & 0       & 10  & 0       & 0       &    0   \\
Diboson Rate              & 0      & 0       & 0   & 11.5    & 0       &    0   \\
Signal Cross Section      & 0      & 0       & 0   & 0       & 0       &    5 \\
HF Fraction in W+jets     &    30.1  & 0       & 0   & 0       & 0       &    0   \\
ISR+FSR+PDF               & 0      & 0       & 0   & 0       & 0       &    3.1 \\ 
QCD Rate                  & 0      & 0       & 0   & 0       & 40      &    0   \\
\end{tabular}
\end{ruledtabular}
\end{table}

\clearpage

\begin{table}[h]
\label{tab:d0systwh1}
\caption{Systematic uncertainties on the signal contributions  for D\O 's
$WH\rightarrow\ell\nu b{\bar{b}}$ single (ST) and double tag (DT) channels,
$WH\rightarrow\tau\nu b\bar{b}$, and $VH \rightarrow \tau\tau b\bar{b}/q\bar{q} \tau\tau$.
Systematic uncertainties are listed by name, see the original 
references for a detailed explanation of their meaning and on how they are derived.  
Systematic uncertainties for $WH$ shown in this table are obtained for $m_H=115$ GeV/c$^2$.
  Uncertainties are
relative, in percent and are symmetric unless otherwise indicated.  }
\vskip 0.2cm
{\centerline{D\O : Single Tag (ST) $WH \rightarrow\ell\nu b\bar{b}$ Analysis}}
\vskip 0.099cm
\begin{ruledtabular}
\begin{tabular}{l c c c c c c c }\\
Contribution  &~WZ/WW~&Wbb/Wcc&Wjj/Wcj&$~~~t\bar{t}~~~$&single top&Multijet& ~~~WH~~~\\ 
\hline                                                                            
Luminosity                &  6    &  6    &  6    &  6    &  6    &  0    &  6    \\ 
Trigger eff.              &  3--5 &  3--5 &  3--5 &  3--5 &  3--5 &  0    &  3--5 \\       
EM ID/Reco eff./resol.    &     5 &     5 &     5 &     5 &     5 &  0    &     5 \\       
Muon ID/Reco eff./resol.  &     5 &     5 &     5 &     5 &     5 &  0    &     5 \\        
Jet ID/Reco eff.          &     3 &     3 &     3 &     3 &     3 &  0    &     3 \\ 
Jet Energy Scale          &     3 &     4 &     3 &     4 &     2 &  0    &     3 \\       
Jet mult./frag./modeling  &     0 &    11 &     9 &     5 &     5 &  0    &     5 \\       
$b$-tagging/taggability   &     4 &     4 &    15 &     4 &     4 &  0    &     4 \\ 
Cross Section             &     6 &     9 &     9 &    10 &    10 &  0    &     6 \\       
Heavy-Flavor K-factor     &  0    &    20 &     0 &  0    &  0    &  0    &  0    \\       
Instrumental-WH           &  0    &  0    &  -6   &  0    &  0    &    26 &  0    \\ 
\end{tabular}                                                                            
\end{ruledtabular}
\vskip 0.5cm                                                                             
{\centerline{D\O : Double Tag (DT) $WH \rightarrow\ell\nu b\bar{b}$ Analysis}}
\vskip 0.099cm
\begin{ruledtabular}
\begin{tabular}{ l c c c c c c c }   \\                                             
Contribution  &~WZ/WW~&Wbb/Wcc&Wjj/Wcj&$~~~t\bar{t}~~~$&single top&Multijet& ~~~WH~~~\\ 
\hline                                                                            
Luminosity                &  6    &  6    &  6    &  6    &  6    &  0    &  6    \\ 
Trigger eff.              &  3--5 &  3--5 &  3--5 &  3--5 &  3--5 &  0    &  3--5 \\       
EM ID/Reco eff./resol.    &     5 &     5 &     5 &     5 &     5 &  0    &     5 \\       
Muon ID/Reco eff./resol.  &     5 &     5 &     5 &     5 &     5 &  0    &     5 \\        
Jet ID/Reco eff.          &     3 &     3 &     3 &     3 &     3 &  0    &     3 \\ 
Jet Energy Scale          &     3 &     4 &     3 &     4 &     2 &  0    &     3 \\       
Jet mult./frag./modeling  &     0 &    11 &     9 &     5 &     5 &  0    &     5 \\       
$b$-tagging/taggability   &     8 &     8 &    25 &     8 &     8 &  0    &     8 \\ 
Cross Section             &     6 &     9 &     9 &    10 &    10 &  0    &     6 \\       
Heavy-Flavor K-factor     &  0    &    20 &     0 &  0    &  0    &  0    &  0    \\       
Instrumental-WH           &  0    &  0    &  -6   &  0    &  0    &    26 &  0    \\ 
\end{tabular}                                                                                           
\end{ruledtabular}
{\centerline{D\O : $WH\rightarrow\tau\nu b{\bar{b}}$ and $VH \rightarrow \tau\tau b\bar{b}/q\bar{q} \tau\tau$ analyses}}
\vskip 0.099cm
\begin{ruledtabular}
\begin{tabular}{ l c c c c }\\

Contribution             &~~~background~~~& ~~~WH~~~&~~background~~~& ~~~VH~~~~\\ 
                         &    &$\tau\nu b{\bar{b}}$          && $\tau\tau b\bar{b}/q\bar{q} \tau\tau$          \\ 
\hline                                                                                                       
Luminosity                &  6            &  6       &  6            &  6       \\ 
Trigger eff.              &  6            &  6       &  3            &  3       \\       
$\mu$  ID                 &    --         &  --      &    5          &  5   \\       
$\tau$ ID                 &    5--6       &  5--6    &     3         &  3   \\       
$\tau$ energy scale       &     3         &     3    &     4         &     4    \\        
Jet ID/Reco eff.          &  2--5         &  2--5    & 2--5          &  2--5    \\ 
Jet Resolution            &     4         &     1    &     4         &     1    \\ 
Jet Energy Scale          &     1         &     1    &     8         &     8    \\       
$b$-tagging/taggability   &     4         &     1    &     4         &     1    \\ 
Cross Section             &   6--18       &     6    &   6--18       &     6    \\       
Heavy-Flavor K-factor     &    30         &  0       &    30         &  0       \\       
Instrumental-WH           &  35--100      &  0       &  35--100      &  0       \\ 
\hline                                                                            
\end{tabular}                                                                                           
\end{ruledtabular}
\end{table}


\begin{table}
\caption{Systematic uncertainties for CDF's $ZH\rightarrow\nu{\bar{\nu}} b{\bar{b}}$ tight (TDT and loose (LDT) double-tag, and single-tag (ST) channel.
Systematic uncertainties are listed by name, see the original references for a detailed explanation of their meaning and on how they are derived.  
Systematic uncertainties for $ZH$ and $WH$ shown in this table are obtained for $m_H=120$ GeV/c$^2$.
Uncertainties are relative, in percent and are symmetric unless otherwise indicated.  }
\label{tab:cdfvvbb1}
\vskip 0.5cm                                                                                                          
{\centerline{CDF: $ZH\rightarrow\nu{\bar{\nu}} b{\bar{b}}$ tight double-tag (TDT) channel}}
\vskip 0.099cm                                                                                                          
\begin{footnotesize}
\begin{ruledtabular}
      \begin{tabular}{lcccccccc}\\
                          & ZH & WH &Multijet& Top Pair & S. Top & Di-boson  & W + h.f.  & Z + h.f. \\\hline
        \multicolumn{9}{l}{\it{Correlated uncertainties}}                                       \\\hline
        Luminosity       & 3.8 & 3.8 &       & 3.8 & 3.8 & 3.8     & 3.8     & 3.8     \\
        Lumi Monitor      & 4.4 & 4.4 &       & 4.4 & 4.4 & 4.4     & 4.4     & 4.4     \\
        Tagging SF        & 8.6 & 8.6 &       & 8.6 & 8.6 & 8.6     & 8.6     & 8.6     \\
      Trigger Eff. (shape)& 1.0 & 1.2 & 1.1 & 0.7 & 1.1 & 1.6     & 1.7     & 1.3     \\
        Lepton Veto       & 2.0 & 2.0 &       & 2.0 & 2.0 &2.0      & 2.0     & 2.0     \\
        PDF Acceptance    & 2.0 & 2.0 &       & 2.0 & 2.0 &2.0      & 2.0     & 2.0     \\
        JES (shape)       & $^{+3.0}_{-3.0}$ 
                                  & $^{+3.5}_{-4.7}$ 
                                          &  $^{-4.0}_{+3.8}$ 
                                                  & $^{+1.1}_{-1.1}$ 
                                                          & $^{+2.4}_{-4.7}$ 
                                                                  & $^{+8.2}_{-6.1}$ 
                                                                             & $^{+7.3}_{-11.8}$  
                                                                                          & $^{+6.5}_{-8.3}$    \\
        ISR               & \multicolumn{2}{c}{$^{+4.4}_{+3.7}$} &       &       &       &           &           &      \\
        FSR               & \multicolumn{2}{c}{$^{+1.8}_{+4.4}$} &       &       &       &           &           &      \\\hline
        \multicolumn{9}{l}{\it{Uncorrelated uncertainties}}                             \\\hline
        Cross-Section     &  5  & 5 &       & 10 & 10 & 11.5    & 40      & 40      \\
        Multijet Norm.  (shape)   &       & & 20.6 &       & &          &           &           \\
      \end{tabular}
\end{ruledtabular}
\end{footnotesize}
\vskip 0.5cm                                                                                                          
{\centerline{CDF: $ZH\rightarrow\nu{\bar{\nu}} b{\bar{b}}$ loose double-tag (LDT) channel}}
\vskip 0.099cm                                                                                                          
\label{tab:cdfvvbb2} 
\begin{footnotesize}
 \begin{ruledtabular}
     \begin{tabular}{lcccccccc}\\
                          & ZH & WH & Multijet & Top Pair & S. Top  & Di-boson  & W + h.f.  & Z + h.f. \\\hline
        \multicolumn{9}{l}{\it{Correlated uncertainties}}                                       \\\hline
        Luminosity       & 3.8  & 3.8  &     & 3.8  & 3.8  & 3.8      & 3.8      & 3.8     \\
        Lumi Monitor      & 4.4  & 4.4  &     & 4.4  & 4.4  & 4.4      & 4.4      & 4.4     \\
        Tagging SF        & 12.4 & 12.4 &     & 12.4 & 12.4 & 12.4     & 12.4     & 12.4     \\
     Trigger Eff. (shape) & 1.2  & 1.3  &1.1& 0.7  & 1.2  & 1.2      & 1.8      & 1.3     \\
        Lepton Veto       & 2.0  & 2.0  &     & 2.0  & 2.0  &2.0       & 2.0      & 2.0     \\
        PDF Acceptance    & 2.0  & 2.0  &     & 2.0  & 2.0  &2.0       & 2.0      & 2.0     \\
        JES (shape)       & $^{+3.7}_{-3.7}$ 
                                   & $^{+4.0}_{-4.0}$ 
                                            & $^{-5.4}_{+5.2}$     
                                                  & $^{+1.1}_{-0.7}$ 
                                                           & $^{+4.2}_{-4.2}$ 
                                                                    & $^{+7.0}_{-7.0}$ 
                                                                                 & $^{+1.3}_{-7.6}$  
                                                                                              & $^{+6.2}_{-7.1}$    \\
        ISR               & \multicolumn{2}{c}{$^{+1.4}_{-2.9}$} &       &       &       &           &           &      \\
        FSR               & \multicolumn{2}{c}{$^{+5.3}_{+2.5}$} &       &       &       &           &           &      \\\hline
        \multicolumn{9}{l}{\it{Uncorrelated uncertainties}}                             \\\hline
        Cross-Section     &  5.0   & 5.0 &      & 10 & 10 & 11.5    & 40      & 40      \\
        Multijet Norm.  (shape)   &       & & 15.6 &       & &          &           &           \\\hline
      \end{tabular}
\end{ruledtabular}
\end{footnotesize}
\vskip 0.5cm                                                                                                          
{\centerline{CDF: $ZH\rightarrow\nu{\bar{\nu}} b{\bar{b}}$ single-tag (ST) channel}}
\vskip 0.099cm                                                                                                          
\label{tab:cdfvvbb3}
\begin{footnotesize}
\begin{ruledtabular}
      \begin{tabular}{lcccccccc}\\
                          & ZH & WH & Multijet & Top Pair & S. Top  & Di-boson  & W + h.f.  & Z + h.f. \\\hline
        \multicolumn{9}{l}{\it{Correlated uncertainties}}                                       \\\hline
        Luminosity       & 3.8  & 3.8  &     & 3.8  & 3.8  & 3.8      & 3.8      & 3.8     \\
        Lumi Monitor      & 4.4  & 4.4  &     & 4.4  & 4.4  & 4.4      & 4.4      & 4.4     \\
        Tagging SF        & 4.3  & 4.3  &     & 4.3  & 4.3  & 4.3      & 4.3      & 4.3     \\
     Trigger Eff. (shape) & 0.9  & 1.1  &1.1& 0.7  & 1.1  & 1.3      & 2.0      & 1.4     \\
        Lepton Veto       & 2.0  & 2.0  &     & 2.0  & 2.0  &2.0       & 2.0      & 2.0     \\
        PDF Acceptance    & 2.0  & 2.0  &     & 2.0  & 2.0  &2.0       & 2.0      & 2.0     \\
        JES (shape)       & $^{+3.8}_{-3.8}$ 
                                   & $^{+3.8}_{-3.8}$ 
                                            & $^{-5.2}_{+5.6}$     
                                                  & $^{+0.7}_{-0.8}$ 
                                                           & $^{+4.6}_{-4.6}$ 
                                                                    & $^{+7.0}_{-5.6}$ 
                                                                                 & $^{+12.4}_{-12.7}$  
                                                                                              & $^{+8.3}_{-8.1}$    \\
        ISR               & \multicolumn{2}{c}{$^{-1.0}_{-1.5}$} &       &       &       &           &           &      \\
        FSR               & \multicolumn{2}{c}{$^{+2.0}_{-0.1}$} &       &       &       &           &           &      \\\hline
        \multicolumn{9}{l}{\it{Uncorrelated uncertainties}}                             \\\hline
        Cross-Section     &  5.0   & 5.0 &      & 10 & 10 & 11.5    & 40      & 40      \\
        Multijet Norm.  (shape)   &       & & 5.5 &       & &          &           &           \\\hline
      \end{tabular}
\end{ruledtabular}
\end{footnotesize}
\end{table}


\begin{table}
\begin{center}
\caption{Systematic uncertainties on the contributions for D\O 's $ZH\rightarrow \nu \nu b{\bar{b}}$ double-tag (DT) channel.
Systematic uncertainties are listed by name, see the original references for a detailed explanation of their meaning and on how they\
 are derived.
Systematic uncertainties for $ZH$, $WH$  shown in this table are obtained for $m_H=115$ GeV/c$^2$.
Uncertainties are
relative, in percent and are symmetric unless otherwise indicated. }
\label{tab:d0vvbb}
\vskip 0.5cm
{\centerline{D\O : Double Tag (DT)~ $ZH \rightarrow \nu\nu b \bar{b}$ Analysis}}
\vskip 0.099cm
\begin{tabular}{| l | c | c | c | c | c | c |}
\hline
Contribution                       &~WZ/ZZ~&~Z+jets~&~W+jets~ &~~~$t\bar{t}$&~~ZH,WH~~\\ \hline
Luminosity                             &  6    &  6    &  6    &  6    &  6      \\
Trigger eff.                           &  5    &  5    &  5    &  5    &  5      \\
Jet Energy Scale                       &  3    &     3 &     3 &     3 &  2      \\
Jet ID/resolution.                     &  2    &     2 &     2 &     2 &  2      \\
$b$-tagging/taggability                  &  6    &  6    &  6    &  6    &  6      \\
Cross Section                          &  6    &  15   &  15   & 18    &  6      \\
Heavy Flavour K-factor                 &  -    &  50   &  50   &  -    &  -      \\
\hline
\end{tabular}
\end{center}
\end{table}

\begin{table}
\caption{Systematic uncertainties on the contributions for CDF's $ZH\rightarrow \ell^+\ell^-b{\bar{b}}$ tight double-tag (TDT) channel.
Systematic uncertainties are listed by name, see the original references for a detailed explanation of their meaning and on how they are derived.  
Systematic uncertainties for $ZH$  shown in this table are obtained for $m_H=115$ GeV/c$^2$.
Uncertainties are relative, in percent and are symmetric unless otherwise indicated. }
\label{tab:cdfllbb1}
\vskip 0.5cm                                                                                                          
{\centerline{CDF: $ZH\rightarrow \ell^+\ell^-b{\bar{b}}$ tight double-tag (TDT) channel}}
\vskip 0.099cm                                                                                                          
\begin{footnotesize}                                                                                         
\begin{ruledtabular}                  
\begin{tabular}{lcccccccc}\\
Contribution   & Fakes & Top  & $WZ$ & $ZZ$ & $Z+b{\bar{b}}$ & $Z+c{\bar{c}}$& $Z+$mistag & $ZH$ \\ \hline
Luminosity ($\sigma_{\mathrm{inel}}(p{\bar{p}})$)          & 0     &    3.8 &    3.8 &    3.8 &    3.8           &    3.8          &    0     &    3.8  \\
Luminosity Monitor        & 0     &    4.4 &    4.4 &    4.4 &    4.4           &    4.4          &    0     &    4.4  \\
Lepton ID    & 0     &    1 &    1 &    1 &    1           &    1          &    0     &    1  \\
Fake Leptons       & 50    &    0 &    0 &    0 &    0           &    0          &    0     &    0  \\
Jet Energy Scale (shape dep.)         & 0     &   
$^{+1.6}_{-1.1}$    & 
0    & 
$^{+1.8}_{-2.7}$    & 
$^{+5.9}_{-6.8}$    & 
$^{+5.9}_{-5.9}$        & 
0   & 
  $^{+2.5}_{+0.7}$   \\  
Mistag Rate      & 0     &    0 &    0 &    0 &    0           &    0          &  $^{+29.2}_{-25.4}$     &    0  \\
$B$-Tag Efficiency      & 0     &   8 &   8 &   8 &   8           &    8         &    0     &   8  \\
$t{\bar{t}}$ Cross Section         & 0     &   10 &    0 &    0 &    0           &    0          &    0     &    0  \\
Diboson Cross Section        & 0     &    0 &   11.5 &   11.5  &    0           &    0          &    0     &    0  \\
Signal Cross Section        & 0     & 0    & 0   & 0    & 0              & 0             & 0        & 5     \\
$\sigma(p{\bar{p}}\rightarrow Z+HF)$      & 0     &    0 &    0 &    0 &   40           &   40          &    0     &    0  \\
ISR (shape dep.)           & 0     & 0    & 0    & 0    & 0              & 0             & 0        &   $^{-1.5}_{+0.7}$     \\
FSR (shape dep.)           & 0     & 0    & 0    & 0    & 0              & 0             & 0        &   $^{-1.7}_{+0.1}$    \\

\end{tabular}
\end{ruledtabular}
\end{footnotesize}
\vskip 0.5cm                                                                                                          
{\centerline{CDF: $ZH\rightarrow \ell^+\ell^-b{\bar{b}}$ loose double-tag (LDT) channel}}
\vskip 0.099cm                                                                                                          
\label{tab:cdfllbb2}
\begin{footnotesize}                                                                                                      
  \begin{ruledtabular}    
\begin{tabular}{lcccccccc}\\
Contribution   & Fakes & Top  & $WZ$ & $ZZ$ & $Z+b{\bar{b}}$ & $Z+c{\bar{c}}$& $Z+$mistag & $ZH$ \\ \hline
Luminosity ($\sigma_{\mathrm{inel}}(p{\bar{p}})$)          & 0     &    3.8 &    3.8 &    3.8 &    3.8           &    3.8          &    0     &    3.8  \\
Luminosity Monitor        & 0     &    4.4 &    4.4 &    4.4 &    4.4           &    4.4          &    0     &    4.4  \\
Lepton ID    & 0     &    1 &    1 &    1 &    1           &    1          &    0     &    1  \\
Fake Leptons       & 50    &    0 &    0 &    0 &    0           &    0          &    0     &    0  \\
Jet Energy Scale (shape dep.)         & 0     &   
$^{+1.3}_{-0.6}$    & 
$^{+3.0}_{-4.4}$    & 
$^{+3.1}_{-3.0}$    & 
$^{+7.4}_{-7.3}$    & 
$^{+6.2}_{-6.0}$        & 
0   & 
  $^{+1.4}_{+0.1}$   \\  
Mistag Rate      & 0     &    0 &    0 &    0 &    0           &    0          &  $^{+40.7}_{-28.7}$     &    0  \\
$B$-Tag Efficiency      & 0     &   11 &   11 &  11 &   11           &    11         &    0     &   11  \\
$t{\bar{t}}$ Cross Section         & 0     &   10 &    0 &    0 &    0           &    0          &    0     &    0  \\
Diboson Cross Section        & 0     &    0 &   11.5 &    11.5 &    0           &    0          &    0     &    0  \\
Signal Cross Section        & 0     & 0    & 0   & 0    & 0              & 0             & 0        & 5     \\
$\sigma(p{\bar{p}}\rightarrow Z+HF)$      & 0     &    0 &    0 &    0 &   40           &   40          &    0     &    0  \\
ISR (shape dep.)           & 0     & 0    & 0    & 0    & 0              & 0             & 0        &   $^{+2.2}_{+1.5}$     \\
FSR (shape dep.)           & 0     & 0    & 0    & 0    & 0              & 0             & 0        &   $^{+1.8}_{-1.1}$    \\
\end{tabular}
\end{ruledtabular}
\end{footnotesize}
\vskip 0.5cm                                                                                                          
{\centerline{CDF: $ZH\rightarrow \ell^+\ell^-b{\bar{b}}$ single-tag (ST) channel}}
\vskip 0.099cm                                                                                                          
\label{tab:cdfllbb3}
\begin{footnotesize}
\begin{ruledtabular}
\begin{tabular}{lcccccccc}\\
Contribution   & Fakes & Top  & $WZ$ & $ZZ$ & $Z+b{\bar{b}}$ & $Z+c{\bar{c}}$& $Z+$mistag & $ZH$ \\ \hline
Luminosity ($\sigma_{\mathrm{inel}}(p{\bar{p}})$)          & 0     &    3.8 &    3.8 &    3.8 &    3.8           &    3.8          & 0        &    3.8  \\
Luminosity Monitor        & 0     &    4.4 &    4.4 &    4.4 &    4.4           &    4.4          & 0        &    4.4  \\
Lepton ID    & 0     &    1 &    1 &    1 &    1           &    1          & 0        &    1  \\
Fake Leptons       & 50    & 0    & 0    & 0    & 0              & 0             & 0        & 0     \\
Jet Energy Scale  (shape dep.)       & 0     & 
  $^{+1.9}_{-2.2}$   & 
  $^{+3.1}_{-4.7}$   & 
  $^{+3.5}_{-5.1}$   & 
  $^{+10.6}_{-9.6}$   & 
  $^{+9.4}_{-9.4}$   & 
  0   & 
  $^{+2.5}_{-2.1}$   \\ 
Mistag Rate      & 0     & 0    & 0    & 0    & 0              & 0             & $^{+12.6}_{-13.2}$    & 0     \\
$B$-Tag Efficiency      & 0     &    4 &    4 &    4 &    4           &   4          & 0        &    4  \\
$t{\bar{t}}$ Cross Section         & 0     &   10 & 0    & 0    & 0              & 0             & 0        & 0     \\
Diboson Cross Section        & 0     & 0    & 11.5   & 11.5    & 0              & 0             & 0        & 0     \\
Signal Cross Section        & 0     & 0    & 0   & 0    & 0              & 0             & 0        & 5     \\
$\sigma(p{\bar{p}}\rightarrow Z+HF)$      & 0     & 0    & 0    & 0    &  40            & 40           & 0        & 0     \\
ISR (shape dep.)           & 0     & 0    & 0    & 0    & 0              & 0             & 0        &   $^{-3.4}_{-3.9}$     \\
FSR (shape dep.)           & 0     & 0    & 0    & 0    & 0              & 0             & 0        &   $^{-0.9}_{-1.7}$     \\
\end{tabular}
\end{ruledtabular}
\end{footnotesize}
\end{table}


\begin{table}
\caption{Systematic uncertainties on the contributions for D\O 's $ZH\rightarrow \ell^+\ell^-b{\bar{b}}$ single-tag (ST) channel.
Systematic uncertainties are listed by name, see the original references for a detailed explanation of their meaning and on how they are derived.  
Systematic uncertainties for $ZH$  shown in this table are obtained for $m_H=115$ GeV/c$^2$.
Uncertainties are relative, in percent and are symmetric unless otherwise indicated. }
\label{tab:d0llbb1}
\vskip 0.8cm                                                                                                          
{\centerline{D\O : Single Tag (ST)~ $ZH \rightarrow \ell\ell b \bar{b}$ Analysis}}
\vskip 0.099cm                                                                                                          
\begin{ruledtabular}
\begin{tabular}{  l  c  c  c  c  c  c  c }                                                                               \\
Contribution & ~~WZ/ZZ~~ &~~Zbb/Zcc~~&~~~Zjj~~~ &~~~~$t\bar{t}$~~~~&  Multijet & ~~~ZH~~~\\ \hline                               
Luminosity                             &  6    &  6    &  6    &  6    &  0    &  6    \\ 
EM ID/Reco eff.                        &  2    &  2    &  2    &  2    &  0    &  2    \\                                      
Muon ID/Reco eff.                      &  2    &  2    &  2    &  2    &  0    &  2    \\                                      
Jet ID/Reco eff.                       &  2    &  2    &  2    &  2    &  0    &  2    \\ 
Jet Energy Scale (shape dep.)          &  5    &  5    &  5    &  5    &  0    &  5    \\                                      
$b$-tagging/taggability                  &  5    &  5    &  5    &  5    &  0    &  5    \\ 
Cross Section                          &  6    & 30    &  6    & 10    &  0    &  6    \\                                      
MC modeling                        &  0    &  4    &  4    &  0    &  0    &  0    \\ 
Instrumental-ZH                        &  0    &  0    &  0    &  0    & 50    &  0    \\ 
\end{tabular}                                                                                                           
\end{ruledtabular} 
\vskip 0.8cm                                                                                                          
{\centerline{D\O : Double Tag (DT)~ $ZH \rightarrow \ell\ell b \bar{b}$ Analysis}}
\vskip 0.099cm                                                                                                          
\begin{ruledtabular}
\begin{tabular}{  l  c  c  c  c  c  c  c }  \\                                                                             

Contribution & ~~WZ/ZZ~~ &~~Zbb/Zcc~~&~~~Zjj~~~ &~~~~$t\bar{t}$~~~~&  Multijet & ~~~ZH~\\ \hline                               
Luminosity                             &  6    &  6    &  6    &  6    &  0    &  6    \\ 
EM ID/Reco eff.                        &  2    &  2    &  2    &  2    &  0    &  2    \\                                      
Muon ID/Reco eff.                      &  2    &  2    &  2    &  2    &  0    &  2    \\                                      
Jet ID/Reco eff.                       &  2    &  2    &  2    &  2    &  0    &  2    \\ 
Jet Energy Scale (shape dep.)          &  5    &  5    &  5    &  5    &  0    &  5    \\                                      
$b$-tagging/taggability                  & 10    & 10    & 10    & 10    &  0    & 10    \\ 
Cross Section                          &  6    & 30    &  6    & 10    &  0    &  6    \\                                      
MC modeling                        &  0    &  4    &  4    &  0    &  0    &  0    \\ 
Instrumental-ZH                        &  0    &  0    &  0    &  0    & 50    &  0    \\
\end{tabular}                                                                                                           
\end{ruledtabular}                                                                                                           
\end{table}

\clearpage

\begin{table}
\begin{footnotesize}
\caption{Systematic uncertainties on the contributions for CDF's
$H\rightarrow W^+W^-\rightarrow\ell^{\pm}\ell^{\prime \mp}$ channels with zero, one, and two or more associated jets.  These channels are 
sensitive to $WH, ZH$ or VBF signals.
Systematic uncertainties are listed by name, see the original references for a detailed explanation of their meaning and on how they are derived. 
Systematic uncertainties for $H$ shown in this table are obtained for $m_H=165$ GeV/c$^2$.
Uncertainties are relative, in percent and are symmetric unless otherwise indicated.  
Uncertainties in bold are correlated across jet bins but not across channels.  Uncertainties in italics
are correlated across jet bins and across appropriate channels.  Monte Carlo statistical uncertainties in each 
bin of each template are considered as independent systematic uncertainties.
}
\label{tab:cdfsystww0}
\vskip 0.3cm                                                                             
{\centerline{CDF: $H\rightarrow W^+W^-\rightarrow\ell^{\pm}\ell^{\prime \mp}$ channels with no associated jet}}
\vskip 0.099cm
\begin{ruledtabular}
\begin{tabular}{lcccccccc}\\
Uncertainty Source         &  $WW$      &  $WZ$         &  $ZZ$   &  $t\bar{t}$  &  DY          &$W\gamma$& $W$+jet(s)&$gg\to H$  \\ \hline 
{Cross Section}        &{\it 6.0}& {\it 6.0}  & {\it 6.0}  &{\it 10.0}&  5.0   & {\it 10.0} &  & {\it 12.0}     \\\hline
Scale (leptons)            &            &               &               &        &              &        &         &  2.5 \\ 
Scale (jets)               &            &               &               &        &              &        &         &  4.6 \\ 
PDF Model (leptons)        &  1.9     &  2.7        &  2.7        &  2.1 &  4.1       &  2.2 &         &  1.5 \\ 
PDF Model (jets)           &            &               &               &        &              &        &         &  0.9 \\ 
Higher-order Diagrams      & {\bf 5.5} & {\bf 10.0} & {\bf 10.0}  & 10.0 & {\bf 5.0}  & {\bf 10.0} &   &        \\ 
Missing Et Modeling        &  1.0     &  1.0        &  1.0        &  1.0 & 21.0       &  1.0 &         &  1.0 \\ 
Conversion Modeling        &            &               &               &        &              & 20.0 &         &        \\ 
Jet Fake Rates   
(Low/High S/B)                  &            &               &               &        &              &        & 21.5/27.7  &        \\
$W\gamma$+jet modeling     &            &              &          &              &              &{\it 4.0}&          &    \\
Lepton ID Efficiencies     &  2.0     &  1.7        &  2.0        &  2.0 &  1.9       &  1.4 &         &  1.9 \\ 
Trigger Efficiencies       &  2.1     &  2.1        &  2.1        &  2.0 &  3.4       &  7.0 &         &  3.3 \\ \hline
{ Luminosity}           &  3.8     &  3.8        &  3.8        &  3.8 &  3.8       &  3.8 &         &  3.8 \\ 
{ Luminosity Monitor}   &  4.4     &  4.4        &  4.4        &  4.4 &  4.4       &  4.4 &         &  4.4 \\ 
\end{tabular}
\end{ruledtabular}
\vskip 0.3cm                                                                             
{\centerline{CDF: $H\rightarrow W^+W^-\rightarrow\ell^{\pm}\ell^{\prime \mp}$ channels with one associated jet}}
\vskip 0.099cm
\begin{ruledtabular}
\begin{tabular}{lccccccccccc} \\
Uncertainty Source         &  $WW$      &  $WZ$         &  $ZZ$  &  $t\bar{t}$   &  DY          & $W\gamma$   & $W$+jet(s) &$gg \to H$&  $WH$  &  $ZH$  &  VBF \\ \hline 
{ Cross Section}        &{\it 6.0}& {\it 6.0}  & {\it 6.0}  &{\it 10.0}&  5.0   & {\it 10.0} &  & {\it 12.0} & {\it 5.0} & {\it 5.0} & {\it 10.0} \\\hline
Scale (leptons)            &            &               &               &        &              &        &              &  2.8 &        &        &        \\ 
Scale (jets)               &            &               &               &        &              &        &              & -5.1 &        &        &        \\ 
PDF Model (leptons)        &  1.9     &  2.7        &  2.7        &  2.1 &  4.1       &  2.2 &              &  1.7 &  1.2 &  0.9 &  2.2 \\ 
PDF Model (jets)           &            &               &               &        &              &        &              & -1.9 &        &        &        \\ 
Higher-order Diagrams      &  {\bf 5.5} & {\bf 10.0} & {\bf 10.0} & 10.0 & {\bf 5.0}  & {\bf 10.0} &        &        & {\it 10.0} & {\it 10.0} & {\it 10.0} \\ 
Missing Et Modeling        &  1.0     &  1.0        &  1.0        &  1.0 & 30.0       &  1.0 &              &  1.0 &  1.0 &  1.0 &  1.0 \\ 
Conversion Modeling        &            &               &               &        &              & 20.0 &              &        &        &        &        \\ 
Jet Fake Rates 
(Low/High S/B)                  &            &               &               &        &              &        & 22.2/31.5       &        &        &        &        \\
$W\gamma$+jet modeling     &            &              &          &              &              &{\it 15.0}&          &  & & &   \\
MC Run Dependence          &  1.8     &               &               &  2.2 &              &  2.2 &              &  2.6 &  2.6 &  1.9 &  2.8 \\ 
Lepton ID Efficiencies     &  2.0     &  2.0        &  2.2        &  1.8 &  2.0       &  2.0 &              &  1.9 &  1.9 &  1.9 &  1.9 \\ 
Trigger Efficiencies       &  2.1     &  2.1        &  2.1        &  2.0 &  3.4       &  7.0 &              &  3.3 &  2.1 &  2.1 &  3.3 \\ \hline
{ Luminosity}           &  3.8     &  3.8        &  3.8        &  3.8 &  3.8       &  3.8 &              &  3.8 &  3.8 &  3.8 &  3.8 \\
{ Luminosity Monitor}   &  4.4     &  4.4        &  4.4        &  4.4 &  4.4       &  4.4 &              &  4.4 &  4.4 &  4.4 &  4.4 \\ 
\end{tabular}
\end{ruledtabular}
\vskip 0.3cm
{\centerline{CDF: $H\rightarrow W^+W^-\rightarrow\ell^{\pm}\ell^{\prime \mp}$ channels with two or more associated jets}
\vskip 0.099cm
}
\label{tab:cdfsystww2}
\begin{ruledtabular}
\begin{tabular}{lccccccccccc}\\
Uncertainty Source         &  $WW$      &  $WZ$         &  $ZZ$  &  $t\bar{t}$  &  DY           &  $W\gamma$    & $W$+jet(s) &$gg\to H$&  $WH$  &  $ZH$  &  VBF         \\ \hline 
{ Cross Section}        &{\it 6.0}& {\it 6.0}  & {\it 6.0}  &{\it 10.0}&  5.0   & {\it 10.0} &  & {\it 12.0} & {\it 5.0} & {\it 5.0} & {\it 10.0} \\\hline
Scale (leptons)            &            &               &               &        &              &               &         &  3.1 &        &        &          \\ 
Scale (jets)               &            &               &               &        &              &               &         & -8.7 &        &        &          \\ 
PDF Model (leptons)        &  1.9     &  2.7        &  2.7        &  2.1 &  4.1       &  2.2        &         &  2.0 &  1.2 &  0.9 &  2.2   \\ 
PDF Model (jets)           &            &               &               &        &              &               &         & -2.8 &        &        &          \\ 
Higher-order Diagrams      & {\bf 10.0} & {\bf 10.0} & {\bf 10.0} & 10.0 & {\bf 10.0} & {\bf 10.0}  &         &        & {\it 10.0} & {\it 10.0} & {\it 10.0}\\
Missing Et Modeling        &  1.0     &  1.0        &  1.0        &  1.0 & 32.0       &  1.0        &         &  1.0 &  1.0 &  1.0 &  1.0   \\ 
Conversion Modeling        &            &               &               &        &              & 20.0        &         &        &        &        &          \\ 
$b$-tag Veto               &            &               &               &  7.0 &              &               &         &        &        &        &          \\ 
Jet Fake Rates             &            &               &               &        &              &               & 27.1  &        &        &        &          \\ 
$W\gamma$+jet modeling     &            &              &          &              &              &{\it 20.0}&          &  & & &   \\
MC Run Dependence          &  1.0     &               &               &  1.0 &              &  1.0        &         &  1.7 &  2.0 &  1.9 &  2.6   \\ 
Lepton ID Efficiencies     &  1.9     &  2.9        &  1.9        &  1.9 &  1.9       &  1.9        &         &  1.9 &  1.9 &  1.9 &  1.9   \\ 
Trigger Efficiencies       &  2.1     &  2.1        &  2.1        &  2.0 &  3.4       &  7.0        &         &  3.3 &  2.1 &  2.1 &  3.3   \\ \hline
{ Luminosity}           &  3.8     &  3.8        &  3.8        &  3.8 &  3.8       &  3.8 &              &  3.8 &  3.8 &  3.8 &  3.8 \\
{ Luminosity Monitor}   &  4.4     &  4.4        &  4.4        &  4.4 &  4.4       &  4.4 &              &  4.4 &  4.4 &  4.4 &  4.4 \\ 
\end{tabular}                                                                                                             
\end{ruledtabular}                                                                                                             
\end{footnotesize}
\end{table}


\begin{table}
\caption{
Systematic uncertainties on the contributions for CDF's
$WH\rightarrow WWW \rightarrow\ell^{\prime \pm}\ell^{\prime \pm}$ channel.
Systematic uncertainties are listed by name, see the original references for a detailed
 explanation of their meaning and on how they are derived. 
Systematic uncertainties for $ZH$, $WH$  shown in this table are obtained for $m_H=165$ GeV/c$^2$.
Uncertainties are relative, in percent and are symmetric unless otherwise indicated.
Uncertainties in bold are correlated across jet bins but not across channels.  Uncertainties in italics
are correlated across jet bins and across appropriate channels.  Monte Carlo statistical uncertainties in each 
bin of each template are considered as independent systematic uncertainties.
 }
\vglue 0.2cm
\label{tab:cdfsystwww}
\vskip 0.8cm                                                                                                          
{\centerline{CDF: $WH \rightarrow WWW \rightarrow\ell^{\pm}\ell^{\prime\pm}$ Analysis.}}
\vskip 0.099cm                                                                                                          
\begin{ruledtabular}
\begin{tabular}{lccccccccc}\\
Uncertainty Source         &  $WW$      &  $WZ$         &  $ZZ$         &$t\bar{t}$&  DY       &  $W\gamma$    & $W$+jet(s) &  $WH$  &     $ZH$     \\ \hline 
{ Cross Section}           &{\it 6.0}   & {\it 6.0}     &  {\it 6.0}    &{\it 10.0}&  5.0      & {\it 10.0}    &            &{\it 5.0}&{\it 5.0}    \\ \hline
PDF Model (leptons)        &  1.9       &  2.7          &  2.7          &  2.1   &  4.1        &  2.2          &            &  1.2   &  0.9         \\ 
PDF Model (jets)           &            &               &               &        &             &               &            &        &              \\ 
Higher-order Diagrams      & {\bf 10.0} & {\bf 10.0}    & {\bf 10.0}    & 10.0   & {\bf 10.0}  & {\bf 10.0}    &            &{\it 10.0}&{\it 10.0}  \\
Missing Et Modeling        &  1.0       &  1.0          &  1.0          &  1.0   & 20.0        &  1.0          &            &  1.0   &  1.0         \\ 
Conversion Modeling        &            &               &               &        &             & 20.0          &            &        &              \\ 
$W\gamma+$jet modeling     &            &               &               &        &             & {\it 16.0}    &            &        &              \\
Jet Fake Rates             &            &               &               &        &             &               & 30.0       &        &              \\ 
Charge Misassignment       &  16.5      &               &               & 16.5   & 16.5        &               &            &        &              \\
MC Run Dependence          &  1.9       &               &               &  1.0   &             &  2.4          &            &        &              \\ 
Lepton ID Efficiencies     &  1.9       &  2.9          &  1.9          &  1.9   &  1.9        &  1.9          &            &  1.9   &  1.9         \\ 
Trigger Efficiencies       &  2.1       &  2.1          &  2.1          &  2.0   &  3.4        &  7.0          &            &  2.1   &  2.1         \\ \hline
{ Luminosity}              &  3.8       &  3.8          &  3.8          &  3.8   &  3.8        &  3.8          &            &  3.8   &  3.8         \\
{ Luminosity Monitor}      &  4.4       &  4.4          &  4.4          &  4.4   &  4.4        &  4.4          &            &  4.4   &  4.4         \\
\end{tabular}                                                                                                             
\end{ruledtabular}                                                                                                             
\end{table}

\begin{table}
\caption{Systematic uncertainties on the contributions for D\O 's
$WH \rightarrow WWW \rightarrow\ell^{\prime \pm}\ell^{\prime \pm}$ channel.
Systematic uncertainties are listed by name, see the original references for a detailed explanation of their meaning and on how they are derived. 
Systematic uncertainties for $WH$ shown in this table are obtained for $m_H=165$ GeV/c$^2$.
Uncertainties are relative, in percent and are symmetric unless otherwise indicated.   }
\label{tab:d0systwww}
\vskip 0.8cm                                                                                                          
{\centerline{D\O : $WH \rightarrow WWW \rightarrow\ell^{\pm}\ell^{\prime\pm}$ Run IIa Analysis.}}
\vskip 0.099cm                                                                                                   
\begin{ruledtabular}       
\begin{tabular}{ l  c  c  c  c  }\\

Contribution                           & ~~WZ/ZZ~~ & Charge flips & ~Multijet~ &~~~~~WH~~~~~  \\ \hline
Luminosity                             & 6 &  0 &  0   & 6 \\
Trigger eff.                           &  5    &  0                     &  0  &  5    \\
Lepton ID/Reco. eff                    & 10    &  0                     &  0  & 10    \\
Cross Section                          &  7    &  0                     &  0  &  6    \\
Normalization                          &  6    &  0                     &  0  &  0    \\ 
Instrumental-$ee$ ($ee$ final state)                                                    
                                       &  0    &  32                    &  15 &  0    \\
Instrumental-$e\mu$ ($e\mu$ final state)                                                  
                                       &  0    &  0                     &  18 &  0    \\
Instrumental-$\mu\mu$ ($\mu\mu$ final state)                                                 
                                       &  0    &  $^{+290}_{-100}$      & 32  &  0    \\ \hline
\end{tabular}
\end{ruledtabular}
\end{table}

\clearpage

\begin{table}
\caption{Systematic uncertainties on the contributions for D\O 's
$H\rightarrow WW \rightarrow\ell^{\pm}\ell^{\prime \mp}$ channels.
Systematic uncertainties are listed by name, see the original references for a detailed explanation of their meaning and on how they are derived. 
Systematic uncertainties shown in this table are obtained for the $m_H=165$ GeV/c$^2$ Higgs selection.
Uncertainties are relative, in percent and are symmetric unless otherwise indicated.   }
\label{tab:d0systww}
\vskip 0.8cm                                                                                                          
{\centerline{D\O : $H\rightarrow WW \rightarrow e^{\pm} e^{ \mp}$ Analysis }}
\vskip 0.099cm      
\begin{ruledtabular}
\begin{tabular}{ l  c  c  c  c  c  c  c}  \\                                                                             

Contribution & Diboson & ~~$Z/\gamma^* \rightarrow \ell\ell$~~&$~~W+jet/\gamma$~~ &~~~~$t\bar{t}~~~~$    & ~~Multijet~~  & ~~~~$H$~~~~      \\ 
\hline                                                                                
Lepton ID                        &  3           &   3           & 3             & 3            & --   &   3            \\ 
Momentum resolution              &  1           &   1           & 1             & 1            & --   &   1            \\ 
Jet Energy Scale                 &  5           &   2           & 0             & 5            & --   &   3            \\ 
Jet identification               &  5           &   1           & 0             & 5            & --   &   3            \\ 
Cross Section/normalization      &  6           &   6           & 13            & 10           &  2   &   6            \\ 
Modeling~~~~~                    &  3           &   5           & 0             & 0            &  0   &   3            \\ 

\end{tabular}                                                                                                           
\end{ruledtabular}                                                                                                           
\vskip 0.8cm                                                                                                          
{\centerline{D\O : $H\rightarrow WW \rightarrow e^{\pm} \mu^{\mp}$ Analysis }}
\vskip 0.099cm      
\begin{ruledtabular}
\begin{tabular}{ l  c  c  c  c  c  c  c} \\                                                                              

Contribution & Diboson & ~~$Z/\gamma^* \rightarrow \ell\ell$~~&$~~W+jet/\gamma$~~ &~~~~$t\bar{t}~~~~$    & ~~Multijet~~  & ~~~~$H$~~~~      \\ 
\hline                                                                                
Lepton ID                        &  1           &   1           & 1             & 1            & --   &   1            \\ 
Momentum resolution              &  1           &   4           & 1             & 0            & --   &   1            \\ 
Jet Energy Scale                 &  2           &   2           & 3             & 5            & --   &   3            \\ 
Jet identification               &  0           &   4           & 1             & 4            & --   &   1            \\ 
Cross Section/normalization      &  6           &   6           & 13            & 10           & 13   &   6            \\ 
Modeling~~~~~                    &  1           &   3           & 0             & 0            &  0   &   2            \\ 

\end{tabular}                                                                                                           
\end{ruledtabular}                                                                                                           
\label{tab:d0systww_mm}
\vskip 0.8cm                                                                                                          
{\centerline{D\O : $H\rightarrow WW \rightarrow \mu^{\pm} \mu^{\mp}$ Analysis }}
\vskip 0.099cm      
\begin{ruledtabular}
\begin{tabular}{ l  c  c  c  c  c  c  c} \\                                                                              

Contribution & Diboson & ~~$Z/\gamma^* \rightarrow \ell\ell$~~&$~~W+jet/\gamma$~~ &~~~~$t\bar{t}~~~~$    & ~~Multijet~~  & ~~~~$H$~~~~      \\ 
\hline                                                                                
trigger                          &  2           &   2           & 2             & 2            & --   &   2            \\ 
Lepton ID                        &  3           &   3           & 3             & 3            & --   &   3            \\ 
Momentum resolution              &  2           &   2           & 2             & 2            & --   &   2            \\ 
Jet Energy Scale                 &  5           &   2           & 0             & 5            & --   &   3            \\ 
Cross Section/normalization      &  6           &   6           & 13            & 10           &  2   &   6            \\ 
Modeling~~~~~                    &  3           &   5           & 0             & 0            &  0   &   3            \\ 

\end{tabular}                                                                                                           
\end{ruledtabular}                                                                                                           
\end{table}

\begin{table}
\caption{Systematic uncertainties on the contributions for D\O 's
  $ t \bar{t} H\rightarrow t \bar{t} b \bar{b}$ channel.
Systematic uncertainties for $ZH$, $WH$  shown in this table are obtained for $m_H=115$ GeV/c$^2$.
Systematic uncertainties are listed by name, see the original references for a detailed explanation of 
their meaning and on how they are derived.  %
Uncertainties are relative, in percent and are symmetric unless otherwise indicated.   
}
\label{tab:d0systtth}
\vskip 0.3cm                                                                                                          
{\centerline{D\O :   $ t \bar{t} H\rightarrow t \bar{t} b \bar{b}$  Analysis}}
\vskip 0.099cm                                                                                                          
\begin{ruledtabular}
\begin{tabular}{lcc}\\
Contribution                             &  ~~~background~~~ & ~~~$t \bar{t} H$~~~    \\  
\hline
Luminosity~~~~                           &  6          &  6    \\ 
lepton ID efficiency                     &  2--3       &  2--3    \\
Event preselection                       &  1          &  1    \\
$W$ +jet modeling                        &   15        & -     \\
Cross Section                            &  10--50     &  10    \\
\end{tabular}
\end{ruledtabular}
\end{table}

\begin{table}
\caption{Systematic uncertainties on the contributions for CDF's
$ H\rightarrow \tau^+\tau^-$ channels.
Systematic uncertainties are listed by name, see the original references for a detailed explanation of 
their meaning and on how they are derived.  %
Systematic uncertainties for $H$ shown in this table are obtained for $m_H=115$ GeV/c$^2$.
Uncertainties are relative, in percent and are symmetric unless otherwise indicated.   The systematic uncertainty 
called ``Normalization''  includes effects of the inelastic $p{\bar{p}}$ cross section, the luminosity monitor acceptance, and the lepton trigger
acceptance. It is considered to be entirely correlated with the luminosity uncertainty.
}
\label{tab:cdfsysttautau}
\vskip 0.3cm                                                                                                          
{\centerline{CDF:   $H \rightarrow \tau^+ \tau^-$ Analysis}}
\vskip 0.099cm                                                                                                          
\begin{ruledtabular}
\begin{tabular}{lcccccccccc}\\
Contribution & $Z/\gamma^* \rightarrow \tau\tau$ & $Z/\gamma^* \rightarrow \ell\ell$ & $t\bar{t}$ & diboson  & jet $\rightarrow \tau$ 
& W+jet & $WH$      & $ZH$  & VBF      & $gg\rightarrow H$    \\  
\hline
Luminosity                    &  3.8   &  3.8   &  3.8   &  3.8  &  -   &    -     &  3.8  &  3.8  &  3.8  &  3.8   \\ 
Luminosity Monitor            &  4.4   &  4.4   &  4.4   &  4.4  &  -   &    -     &  4.4  &  4.4  &  4.4  &  4.4   \\ 
$e,\mu$ Trigger               &  1   &  1   &  1   &  1  &  -   &    -     &  1  &  1  &  1  &  1   \\
$\tau$  Trigger               &  3   &  3   &  3   &  3  &  -   &    -     &  3  &  3  &  3  &  3   \\
$e,\mu,\tau$ ID               &  3   &  3   &  3   &  3  &  -   &    -     &  3  &  3  &  3  &  3   \\
PDF Uncertainty               &  1   &  1   &  1   &  1  &  -   &    -     &  1  &  1  &  1  &  1   \\
ISR/FSR                       &  -   &  -   &  -   &  -  &  -   &    -     & 2/0 & 1/1 & 3/1 & 12/1 \\
JES (shape)                   & 16   & 13   &  2   & 10  &  -   &    -     &  3  &  3  &  4  & 14   \\
Cross Section or Norm.        &  2  &  2   & 10   & 11.5  &  -   &   15     &  5  &  5  & 10  & 10   \\
MC model                      & 20   & 10   &  -   &  -  &  -   &    -     &  -  &  -  &  -  &  -   \\
\end{tabular}
\end{ruledtabular}
\end{table}

\begin{table}
\label{tab:cdfjjbbsyst}
\caption[]{Systematic uncertainties summary for CDF's $WH+ZH\rightarrow jjbb$ channel.
Systematic uncertainties are listed by name, see the original references for a detailed explanation of 
their meaning and on how they are derived. 
Uncertainties with provided shape systematics are labeled with ``s''.
Systematic uncertainties for $H$ shown in this table are obtained for $m_H=115$ GeV/c$^2$.
Uncertainties are relative, in percent and are symmetric unless otherwise indicated.   
The cross section uncertainties are uncorrelated with each other (except for single top and $t{\bar{t}}$, which are
treated as correlated).  The QCD uncertainty is also uncorrelated with other channels' QCD rate uncertainties.
}
\vskip 0.3cm                                                                                                          
{\centerline{CDF: $WH+ZH\rightarrow jjbb$ Analysis}}
\vskip 0.099cm
\begin{ruledtabular}
\begin{tabular}{lccccccccc}\\
               &  QCD   &  $t{\bar{t}}$   &  $Wb{\bar{b}}$   &  $WZ$   &  Single Top &  $Z$+jets   &  $WH$   &  $ZH$   \\\hline
 Interpolation & 0s     & --              & --               & --      &  --         &  --         &   --    &   -- \\
 MC Modeling   & 0s     & --              & --               & --      &  --         &  --         &   18s   &  16s \\
 Cross Section & 10     & 10              & 30               & 6       &  10         & 30          & 5       &  5 \\
\hline
\end{tabular}
\end{ruledtabular}
\end{table}

\begin{table}
\caption{Systematic uncertainties on the contributions for D\O 's
$ H\rightarrow \gamma \gamma$ channels.
Systematic uncertainties for $ZH$, $WH$  shown in this table are obtained for $m_H=115$ GeV/c$^2$.
Systematic uncertainties are listed by name, see the original references for a detailed explanation of 
their meaning and on how they are derived.  %
Uncertainties are relative, in percent and are symmetric unless otherwise indicated.   
}
\label{tab:d0systgg}
\vskip 0.3cm                                                                                                          
{\centerline{D\O :   $H \rightarrow \gamma \gamma$ Analysis}}
\vskip 0.099cm       
                                                                                                   
\begin{ruledtabular}
\begin{tabular}{lcc}\\
Contribution &  ~~~background~~~  & ~~~$H$~~~    \\  
\hline
Luminosity~~~~                            &  6     &  6    \\ 
Acceptance                                &  --    &   2   \\
electron ID efficiency                    &  2     &  2    \\
electron track-match inefficiency         & 10--20 & -     \\
Photon ID efficiency                      &  7     &   7     \\
Photon energy scale                       &  --    &   2     \\
Acceptance                                &  --    &  2    \\
$\gamma$-jet and jet-jet fakes          &  26    &  --    \\
Cross Section ($Z$)                       &  4     &  6    \\
Background subtraction                    &  8--14 &  -    \\
\hline
\end{tabular}
\end{ruledtabular}
\end{table}

\clearpage

 \begin{figure}[t]
 \begin{centering}
 \includegraphics[width=14.0cm]{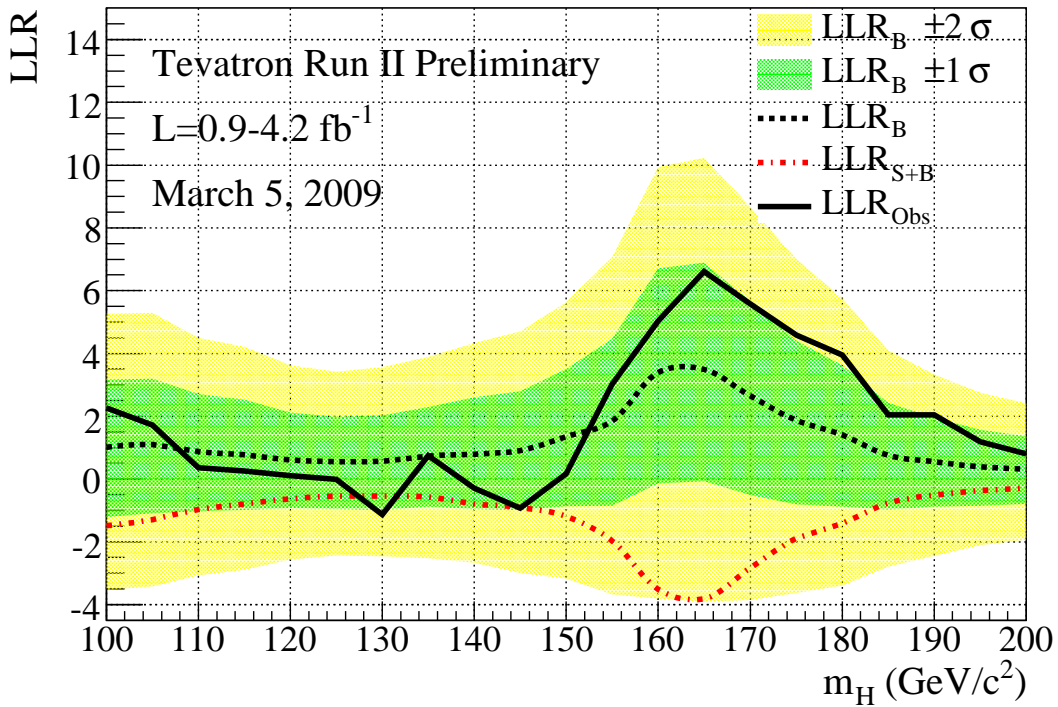}
 \caption{
 \label{fig:comboLLR} { 
Distributions of the log-likelihood ratio (LLR) as a function of Higgs mass obtained with the $CL_s$ method for the combination of all CDF and D\O\ analyses. 
}}
 \end{centering}
 \end{figure}

\vspace*{1cm}
\section{Combined Results} 

Before extracting the combined limits we study the distributions 
of the 
log-likelihood ratio (LLR) for different hypotheses,
to check the expected
sensitivity across the mass range tested.
Figure~\ref{fig:comboLLR}
displays the LLR distributions
for the combined
analyses as a function of $m_{H}$. Included are the results for the
background-only hypothesis (LLR$_{b}$), the signal-plus-background
hypothesis (LLR$_{s+b}$), and for the data (LLR$_{obs}$).  The
shaded bands represent the 1 and 2 standard deviation ($\sigma$)
departures for LLR$_{b}$. 

These
distributions can be interpreted as follows:
The separation between LLR$_{b}$ and LLR$_{s+b}$ provides a
measure of the discriminating power of the search; 
 the size of the 1- and 2-$\sigma$ LLR$_{b}$ bands
provides an estimate of how sensitive the
analysis is to a signal-plus-background-like fluctuation in data, taking account of
the systematic uncertainties;
the value of LLR$_{obs}$ relative to LLR$_{s+b}$ and LLR$_{b}$
indicates whether the data distribution appears to be more signal-plus-background-like
(i.e. closer to the LLR$_{s+b}$ distribution, which is negative by
construction)
or background-like; the significance of any departures
of LLR$_{obs}$ from LLR$_{b}$ can be evaluated by the width of the
LLR$_{b}$ bands.

Using the combination procedures outlined in Section III, we extract limits on
SM Higgs boson production $\sigma \times B(H\rightarrow X)$ in
\pp~collisions at $\sqrt{s}=1.96$~TeV for $m_H=100-200$ GeV/c$^2$.
To facilitate comparisons with the standard model and to accommodate analyses with
different degrees of sensitivity, we present our results in terms of
the ratio of obtained limits  to  cross section in the SM, as a function of
Higgs boson mass, for test masses for which
both experiments have performed dedicated searches in different channels.
A value of the combined limit ratio which is less than or equal to one would indicate that
that particular Higgs boson mass is excluded at the
95\% C.L.

The combinations of results of each single experiment, as used in this Tevatron combination,
yield the following ratios of 95\% C.L. observed (expected) limits to the SM 
cross section: 
3.6~(3.2) for CDF and 3.7~(3.9) for D\O\ at $m_{H}=115$~GeV/c$^2$, and 
1.5~(1.6) for CDF and 1.3~(1.8) for D\O\ at $m_{H}=165$~GeV/c$^2$.

The ratios of the 95\% C.L. expected and observed limit to the SM
cross section are shown in Figure~\ref{fig:comboRatio} for the
combined CDF and D\O\ analyses.  The observed and median expected
ratios are listed for the tested Higgs boson masses in
Table~\ref{tab:ratios} for $m_{H} \leq 150$~GeV/c$^2$, and in
Table~\ref{tab:ratios-3} for $m_{H} \geq 155$~GeV/c$^2$, 
as obtained by the Bayesian and the $CL_S$ methods. 
In the following summary we quote only the limits obtained with the 
Bayesian method since they are slightly more conservative (based on the 
expected limits) for the quoted values, but all the equivalent numbers for 
the $CL_S$ method can be retrieved from the tables.
We obtain the observed
(expected) values of 2.5 (2.4) at $m_{H}=115$~GeV/c$^2$, 0.99 (1.1) at
$m_{H}=160$~GeV/c$^2$, 0.86 (1.1) at $m_{H}=165$~GeV/c$^2$, and 0.99
(1.4) at $m_{H}=170$~GeV/c$^2$.  We exclude at the 95\% C.L. the
production of a standard model Higgs boson with mass between 160 and
170 GeV/c$^2$.  This result is obtained with both Bayesian and $CL_S$
calculations.
\begin{figure}[hb]
\begin{centering}
\includegraphics[width=16.5cm]{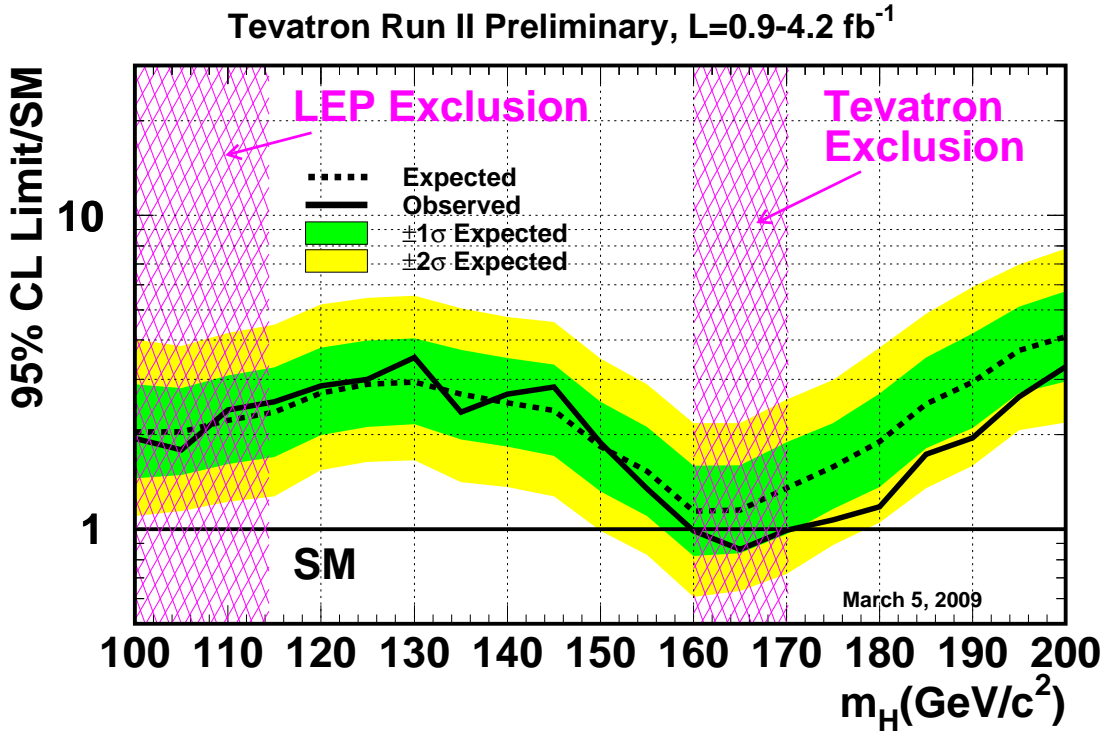}
\caption{
\label{fig:comboRatio}
Observed and expected (median, for the background-only hypothesis)
  95\% C.L. upper limits on the ratios to the
SM cross section, 
as functions of the Higgs boson mass 
for the combined CDF and D\O\ analyses.
The limits are expressed as a multiple of the SM prediction
for test masses (every 5 GeV/$c^2$)
for which both experiments have performed dedicated
searches in different channels.
The points are joined by straight lines 
for better readability.
  The bands indicate the
68\% and 95\% probability regions where the limits can
fluctuate, in the absence of signal. 
The limits displayed in this figure
are obtained with the Bayesian calculation.
}
\end{centering}
\end{figure}

\begin{table}[ht]
\caption{\label{tab:ratios} Ratios of median expected and observed 95\% C.L.
limit to the SM cross section for the combined CDF and D\O\ analyses as a function
of the Higgs boson mass in GeV/c$^2$, obtained with the Bayesian and with the $CL_S$ method.}
\begin{ruledtabular}
\begin{tabular}{lccccccccccc}\\
Bayesian       & 100 &   105  & 110 & 115 & 120 & 125 & 130 & 135 & 140 & 145 & 150\\ \hline 
Expected       & 2.0 &   2.0  & 2.2 & 2.4 & 2.7 & 2.9 & 2.9 & 2.7 & 2.5 & 2.4 & 1.8\\
Observed       & 1.9 &   1.8  & 2.4 & 2.5 & 2.8 & 3.0 & 3.5 & 2.4 & 2.7 & 2.8 & 1.9\\

\hline
\hline\\
$CL_S$         & 100 &    105 & 110 & 115 & 120 & 125 & 130 & 135 & 140 & 145 & 150\\ \hline 
Expected       & 1.9 &  1.9   & 2.1 & 2.4 & 2.6 & 2.7 & 2.9 & 2.7 & 2.5 & 2.2 & 1.8\\
Observed       & 1.7 &  1.7   & 2.2 & 2.6 & 2.8 & 2.9 & 4.0 & 2.6 & 3.1 & 2.8 & 2.0\\
\end{tabular}
\end{ruledtabular}
\end{table}

\begin{table}[ht]
\caption{\label{tab:ratios-3} Ratios of median expected and observed 95\% C.L.
limit to the SM cross section for the combined CDF and D\O\ analyses as a function
of the Higgs boson mass in GeV/c$^2$, obtained with the Bayesian and with the $CL_S$ method.}
\begin{ruledtabular}
\begin{tabular}{lccccccccccc}
Bayesian             &   155  & 160 & 165 & 170 & 175 & 180 & 185 & 190 & 195 & 200\\ \hline 
Expected             &   1.5  & 1.1 & 1.1 & 1.4 & 1.6 & 1.9 & 2.2 & 2.7 & 3.5 & 4.2\\
Observed             &   1.4  & 0.99& 0.86& 0.99& 1.1 & 1.2 & 1.7 & 2.0 & 2.6 & 3.3\\
\hline
\hline\\
$CL_S$         &         155  & 160 & 165 & 170 & 175 & 180 & 185 & 190 & 195 & 200\\ \hline 
Expected             &   1.5  & 1.1 & 1.1 & 1.3 & 1.6 & 1.8 & 2.5 & 3.0 & 3.5 & 3.9\\
Observed             &   1.3  & 0.95& 0.81& 0.92& 1.1 & 1.3 & 1.9 & 2.0 & 2.8 & 3.3\\
\end{tabular}
\end{ruledtabular}
\end{table}

We also show in Figure~\ref{fig:comboLLR-2} the 1-$CL_S$ distribution as a function of 
the Higgs boson mass, at high mass ($\geq 150 $ GeV/$c^2$) which is directly interpreted
as the level of exclusion at 95\% C.L. of our search. Note that this figure is obtained
using the $CL_S$ method. The 90\% C.L. line is also shown on the figure.
We  provide the Log-likelihood ratio (LLR) values for our combined Higgs boson search,
as obtained using the $CL_S$ method in table XX.

In summary, we have combined all available CDF and D\O\ results on SM Higgs search,
based on luminosities ranging from 0.9 to 4.2 fb$^{-1}$.
Compared to our previous combination, new channels have been added and
most previously used channels have been
reanalyzed to gain sensitivity. We use the latest parton distribution
functions and $gg \rightarrow H$ theoretical cross sections when
comparing our limits to the SM predictions at high mass.  

The 95\% C.L. upper limits
on Higgs boson production are a factor of 2.5~(0.86) times the SM
cross section for a Higgs boson mass of $m_{H}=$115~(165)~GeV/c$^2$.
Based on simulation, the corresponding median expected upper limits
are 2.4 (1.1). Standard Model branching ratios, calculated as
functions of the Higgs boson mass, are assumed. These results extend
significantly the individual limits of each collaboration and our
previous combination. The mass range excluded at 95\% C.L. for a SM
Higgs has been extended to  $160<m_{H}<170$~GeV/c$^{2}$.
The sensitivity of our combined search is expected to grow substantially in
the near future with the additional luminosity already recorded at the Tevatron
and not yet analyzed, and with additional improvements of our analysis techniques
which will be propagated in the current and future analyses.

 \begin{figure}[t]
 \begin{centering}
 \includegraphics[width=14.0cm]{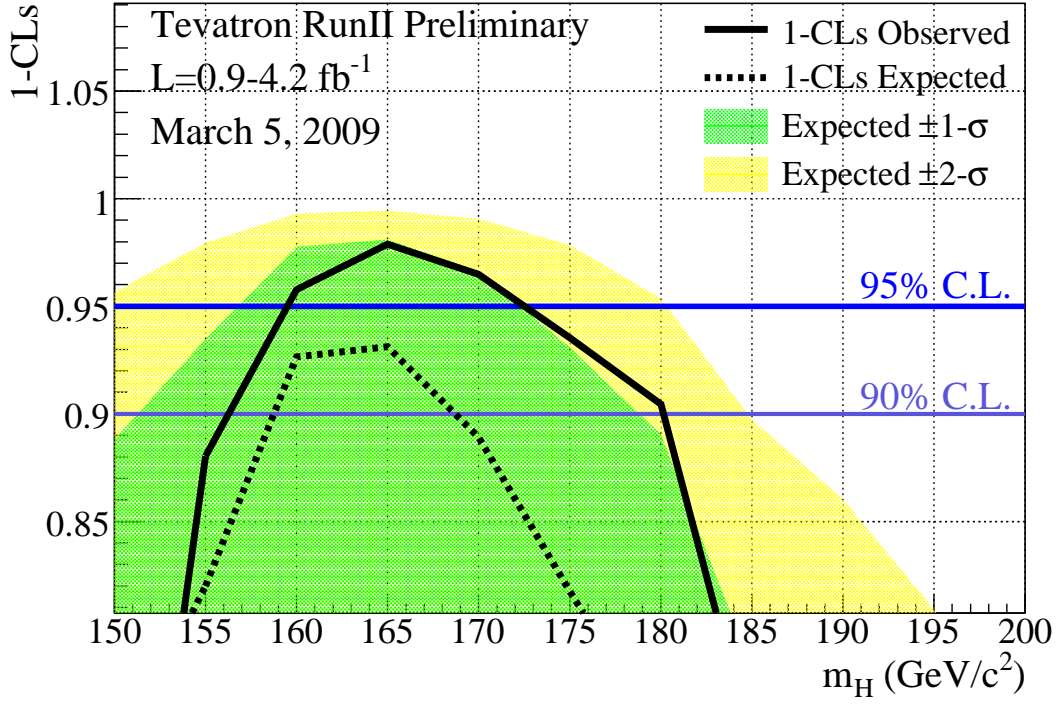}
 \caption{
 \label{fig:comboLLR-2}
 Distributions of 1-$CL_S$ as a function of the Higgs boson mass 
(in steps of 5 GeV/c$^2$), as obtained with $CL_S$ method.
 for the combination of the
 CDF and D\O\ analyses. }
 \end{centering}
 \end{figure}


\begin{table}[htpb]
\caption{\label{tab:llrVals} Log-likelihood ratio (LLR) values for the
combined CDF + \Dzero Higgs boson search obtained using the CL$_{S}$
method.}
\begin{ruledtabular}
\begin{tabular}{lccccccc}
\\ 
$m_{H}$ (GeV/c$^2$) &  LLR$_{obs}$ & LLR$_{S+B}$ &  LLR$_{B}^{-2sigma}$ & LLR$_{B}^{-1sigma}$ & LLR$_{B}$ &  LLR$_{B}^{+1sigma}$ & LLR$_{B}^{+2sigma}$ \\ 
\hline
100 & 2.26 & -1.49 & 5.27 & 3.17 & 1.01 & -1.21 & -3.53 \\ 
105 & 1.72 & -1.29 & 5.29 & 3.19 & 1.09 & -1.09 & -3.43 \\ 
110 & 0.35 & -0.97 & 4.49 & 2.71 & 0.87 & -1.03 & -3.07 \\ 
115 & 0.26 & -0.79 & 4.21 & 2.53 & 0.77 & -0.99 & -2.89 \\ 
120 & 0.11 & -0.63 & 3.63 & 2.11 & 0.61 & -0.93 & -2.57 \\ 
125 & -0.01 & -0.55 & 3.41 & 1.99 & 0.55 & -0.95 & -2.43 \\ 
130 & -1.13 & -0.55 & 3.57 & 2.03 & 0.57 & -0.99 & -2.47 \\ 
135 & 0.75 & -0.57 & 3.89 & 2.29 & 0.73 & -0.89 & -2.53 \\ 
140 & -0.29 & -0.79 & 4.33 & 2.61 & 0.79 & -0.95 & -2.67 \\ 
145 & -0.93 & -0.91 & 4.71 & 2.79 & 0.91 & -0.99 & -2.99 \\ 
150 & 0.16 & -1.19 & 5.63 & 3.51 & 1.35 & -0.87 & -3.17 \\ 
155 & 3.02 & -1.97 & 7.07 & 4.47 & 1.85 & -0.85 & -3.67 \\ 
160 & 5.02 & -3.51 & 9.93 & 6.71 & 3.39 & -0.15 & -3.81 \\ 
165 & 6.60 & -3.81 & 10.23 & 6.89 & 3.49 & -0.07 & -3.91 \\ 
170 & 5.58 & -2.83 & 8.67 & 5.67 & 2.65 & -0.51 & -3.85 \\ 
175 & 4.59 & -1.89 & 7.01 & 4.41 & 1.87 & -0.81 & -3.63 \\ 
180 & 3.95 & -1.41 & 5.75 & 3.63 & 1.41 & -0.89 & -3.39 \\ 
185 & 2.04 & -0.77 & 4.09 & 2.43 & 0.75 & -0.97 & -2.79 \\ 
190 & 2.04 & -0.51 & 3.33 & 1.95 & 0.55 & -0.89 & -2.45 \\ 
195 & 1.19 & -0.37 & 2.75 & 1.57 & 0.39 & -0.85 & -2.11 \\ 
200 & 0.80 & -0.29 & 2.41 & 1.37 & 0.31 & -0.79 & -1.93 \\ 
\hline
\end{tabular}
\end{ruledtabular}
\end{table}

\end{document}